\documentclass[10pt]{article}
\pdfoutput=1
\usepackage{jheppub}
\usepackage[T1]{fontenc}
\usepackage{cancel,tensor,enumitem,fix-cm}
\usepackage{epsfig}
\usepackage{graphicx}
\usepackage[english]{babel}
\usepackage{mathtools}
\usepackage{amsmath}
\usepackage{relsize}
\usepackage{slashed}
\usepackage{textcomp}
\usepackage{multirow}
\usepackage{booktabs}
\usepackage{tikz}
\usepackage{overpic}
\usepackage{bm}
\usepackage{physics}
\usepackage[lofdepth,lotdepth]{subfig}
\usepackage{empheq}
\usepackage{hyperref}

\newcommand{\be}{\begin{equation}}
\newcommand{\ee}{\end{equation}}
\newcommand{\bea}{\begin{eqnarray}}
\newcommand{\eea}{\end{eqnarray}}

\newcommand{\diff}{{\rm d}}

\def\({\left(}
\def\){\right)}
\def\[{\left[}
\def\]{\right]}
\def\cO{{\cal O}}

\subheader{\begin{flushright}
\end{flushright}}

\title{Mutual information from modular flow in CFTs}

\author[a]{Cesar A. Ag\'on,}
\author[b]{Horacio Casini,}
\author[a]{Umut G\"ursoy}
\author[a]{and Guim Planella Planas}
\affiliation[a]{Institute for Theoretical Physics, Utrecht University, 3584 CE Utrecht, The Netherlands}
\affiliation[b]{Instituto Balseiro, Centro At\'omico Bariloche, 8400-S.C de Bariloche, R\'io Negro, Argentina}

\emailAdd{c.a.agonquintero@uu.nl}
\emailAdd{horaciocasini@gmail.com}
\emailAdd{u.gursoy@uu.nl}
\emailAdd{g.planellaiplanas1@uu.nl}

\abstract{
The operator product expansion (OPE) of twist operators in the replica trick framework enables a long-distance expansion of the mutual information (MI) in conformal field theories (CFTs). In this expansion, the terms are labeled by primary operators, as contributions from descendant operators can be resummed into conformal blocks. However, for the MI, the expansion involves primaries from the multi-replica theory, which includes far more operators than those in the original theory. In this work, we develop a method to resum this series, yielding an expansion in terms of the primaries of the original theory, specifically restricted to the two-copy sector. This is achieved by expressing the twist operators in a non-local manner across different replicas and using a modular flow representation to obtain the $n\to 1$ limit of the R\'enyi index. We explicitly compute the resulting ``enhanced conformal blocks,'' which, surprisingly, provide excellent approximations to the MI of generalized free fields across the full range of cross ratios. Remarkably, this approximation appears to be exact in the limit of large spacetime dimensions.

}

\begin{document}
\maketitle
\flushbottom

\section{Introduction}

Mutual information (MI) gives a well-defined information-theoretic measure of correlations between algebras associated to regions in quantum field theory (QFT). It is also a universal regulator of entanglement entropy, allowing it to isolate its cutoff-independent parts. 

There are strong indications to the effect that the vacuum  MI for arbitrary regions contains all the relevant information determining the underlying QFT model. For instance, entanglement entropies are famously known to contain information about the central charge in two-dimensional critical systems or trace anomalies for conformal field theories in higher dimensions.  It is an important goal in this line of research to develop a dictionary between information quantities and more conventional QFT ones. 
There are two aspects in this investigation. One is to develop computational methods to evaluate mutual information in a given model. Another one is that given the knowledge of the mutual information as a function of the regions, one would also like to develop inversion formulas. In other words, we would like to use mutual information from various regions to extract the relevant QFT data. Indeed, in a recent work, the authors used this philosophy to rule out an extensive model of mutual information \cite{Casini:2008wt} from the inconsistencies found in the CFT data extracted from the model \cite{Agon:2021zvp}.  

In the computational front, important progress has been made for the MI of CFT's in the large separation regime. The calculation starts by associating the R\'enyi entropies of index $n$ to the expectation values of twist operators for the $Z_n$ symmetry in the replicated theory CFT$^{\otimes n}$ \cite{Calabrese:2004eu}. In the large separation regime, twist operators can be expanded in an OPE series, giving the connection with ordinary field theory correlators \cite{Headrick:2010zt,Calabrese:2010he, Cardy:2013nua}. It is plain that this series expansion of the MI contains the full information of the CFT spectrum.  
For the special case of spheres, conformal symmetry allows in principle for a complete computation of the series coefficient. The first term in the series is well known to be produced by the lowest dimensional primary in the two-copy sector of the replicated model \cite{Agon:2015ftl}. Contributions of descendant fields can be included by an expansion in terms of conformal blocks \cite{Chen:2017hbk, Agon:2015twa, Chen:2016mya}. A simplification on how to make the analytic continuation to $n\rightarrow 1$, corresponding to the MI, has been achieved by associating the replica index with modular flows \cite{Casini:2021raa}. This simplification has allowed a derivation of the first terms of the expansion for field with spin \cite{Casini:2021raa}. The relation of the three-partite information and higher information measures with OPE coefficients has been explored \cite{Agon:2021lus,Agon:2022efa,Agon:2024zae}.         

While the OPE expansion allows for a calculation of MI order by order in the large distance regime, it poses important technical challenges. Calculations based on the replica trick work within the theory QFT$^{\otimes n}$ as an intermediate step. Such theory has an operator content that differs from the one of the actual QFT, with a large proliferation of primaries with respect to the original theory, even for the two-copy sector.  
 Typically the formulas obtained in the $n\to 1$ limit, depend explicitly on contributions from these multi-copy primary operators. In this paper, we develop a method to re-sum contributions from multi-copy primaries into single functions labeled by original theory primaries. The main idea is to expand the twist operators into multiple integrals of field operators of different copies. The relevant information on the kernels of this expansion can be obtained in the $n\rightarrow 1$ limit by using an analytic continuation in $n$ based on the fact that replica twist operators implement a modular flow in imaginary parameter. We obtain this resumation formula explicitly for the contribution of all two copy primaries (and their descendant fields) that arise from a single primary in the original model, formula (\ref{int-rep-2}) below. 
 
An important aspect of the MI between two spheres in a CFT is that the short distance expansion contains the renormalization group charges: the logarithmic term in the MI is proportional to the trace anomaly for even dimensions and the constant term is the $F$ charge in odd dimensions. In this sense, a more complete knowledge of the MI long-distance expansion holds the key for expressing these renormalization group charges in terms of the CFT data. In other words, this is the unknown form of the Cardy-like formulas for higher dimensional CFT's. Of course, this program would involve an understanding of the re-summation between long-distance and short-distance expansions, that is missing at present. In this sense, it is interesting to note that our formula that re-sums only some specific part of the expansion, does not display an area law at short distances, as is the case of MI for ordinary CFT's,  but rather a volume law. This is to be attributed to missing contributions of larger number of replicas, and the re-summation over different primaries. However, certain non-conventional CFTs, the generalized free fields (GFF), do display volume laws at short distances \cite{Benedetti:2022aiw}. It is an interesting fact that our formulas approximate the MI for generalized free fields quite accurately, especially for larger space-time dimensions, in all the range of distances.        

 An outline of the paper is the following. In the next section, we introduce the long-distance expansion of MI obtained from the OPE expansion of twist operators and the technique for the analytic extension in the R\'enyi index $n$ using modular flows. In section \ref{MI-Local}, we expand the twist in local multi-copy primaries integrated on the domain of dependence of the sphere.  This allows us to recover an expansion in terms of conformal blocks containing contributions from all descendants fields. In section \ref{MI-Multi-local}, we expand in non-local kernels allowing for different positions for the multi-copy primaries. This gives us our main formulas that re-sum contributions from all the two-copy primaries arising from one primary in the single-copy theory. We perform the same calculation in Euclidean signature in section \ref{Euclidean}. We compare with the real-time calculation pointing out the role of the shadow contributions in the Euclidean calculation. We end with a discussion on technical issues and prospects for future work. Some technical results are relegated to appendices.    

\section{Mutual information: Generalities}
We are interested in computing universal contributions to the ground state mutual information between two disjoint spheres in arbitrary dimensions. In conformal field theories, and, in the long separation regime there has been significant progress in computing the leading contribution to the mutual information of CFT primary operators. A remarkable set of results can be found in \cite{Casini:2021raa}. 

The mutual information between two regions $A$ and $B$ can be expressed in terms of the entanglement entropies (von Neumann entropies) of the individual regions as 
\bea\label{mutual-EE}
I(A:B)=S(A)+S(B)-S(A\cup B)\,.
\eea
One standard way of obtaining the entanglement entropy (EE) is via the so-called replica trick method, which entails the evaluation of the R\'enyi entropies for integer $n$ as
 \bea\label{renyis}
S^{(n)}(A)=\frac{1}{1-n}\log\left[\frac{Z({\cal C}^{(n)}_A)}{Z^n({\cal C})}\right]\,,
\eea
so that analytically continuing for real $n$ and taking the $n \to 1$ limit yields the EE of region $A$,
\bea\label{replica-trick}
S(A)=\lim_{n\rightarrow 1} S^{(n)}(A)\,.
\eea
In (\ref{renyis}), ${\cal C}^{(n)}_A$ represents the replica manifold obtained from  $n$ copies of the original spacetime geometry ${\cal C}$ cut along $A$, after suitably identifying one of the two open boundaries along $A$ of the $i$-th copy with another one of the $(i+1)$-th copy, and making $n+1\equiv 1$. We denote $Z(X)$ to the partition function of the theory on the manifold $X$.

The identification process required to define the manifold ${\cal C}^{(n)}_A$ can be implemented by introducing a twist operator $\Sigma^{(n)}_A$ with support on the entangling region  \cite{Calabrese:2004eu,Calabrese:2005zw,Hung:2014npa}. This operator identifies the operators on different consecutive replicas in the region $A$, namely, $\phi_i(x)\to \phi_{i+1}(x)$ when $x\in A$, while acting as the identity when $x\notin A$. $\Sigma^{(n)}_A$ can be normalized with respect to its expectation value, as\footnote{Notice that this definition implies $\langle \tilde{\Sigma}_{A}^{(n)} \rangle=0$.}
\bea
\label{Sigma-A}
\Sigma_{A}^{(n)}= \langle \Sigma_{A}^{(n)} \rangle(1+\tilde{\Sigma}_{A}^{(n)})\, \quad { \rm with} \quad  \langle \Sigma_{A}^{(n)} \rangle= \frac{Z({\cal C}^{(n)}_A)}{Z^n({\cal C})}\,.
\eea
Putting this into equations (\ref{renyis}), (\ref{replica-trick}), and (\ref{mutual-EE}) one gets 
\bea\label{mutual-corr}
I(A:B)=\lim_{n\to 1}\frac{1}{n-1}\langle \tilde{\Sigma}_{A}^{(n)}\tilde{\Sigma}_{B}^{(n)} \rangle \,.
\eea
This formula tells us that the mutual information is just the $n\to 1$ limit of a correlator in the replica theory between operators localized in regions $A$ and $B$ respectively.  It also tells us that understanding the twist operators is essential in the study of mutual information in quantum field theories. 

\subsection{Two-copy sector}
A typical way of studying a given operator in QFT is to compute its correlation function with an arbitrary number of local operators. In the context of the replicated theory QFT$^{\otimes n}$, the simplest contributions to the mutual information come from operators with non-trivial support on two different copies, say ${\cal O}^l(x_1){\cal O}^k(x_2)$ where the indices $1\leq l\neq k\leq n$ label the different copies. We call the set of operators of this form the two-copy sector of the replicated theory. One copy operators do not contribute to the mutual information.

In \cite{Casini:2021raa}, a formula for the vacuum expectation value of the twist operator in the presence of two replica operators was presented. We would like to take that formula, namely \bea\label{twist-correlator}
\frac{\langle \Omega |\Sigma^{(n)}_A {\cal O}^{l}(x_1) {\cal O}^{k}(x_2)|\Omega\rangle}{\langle \Omega |\Sigma^{(n)}_A|\Omega\rangle}=\left\{\begin{array}{ll}
\frac{\tr\left\{\rho_A^{n}\,{\cal O}(x_1) \rho_A^{(l-k)} {\cal O}(x_2)\rho_A^{-(l-k)}\right\}}{\tr\rho^n_A}\,, & {\rm for}\quad l>k\\
\frac{\tr\left\{\rho_A^{n}\,{\cal O}(x_2) \rho_A^{(k-l)} {\cal O}(x_1)\rho_A^{-(k-l)}\right\}}{\tr\rho^n_A}\,, &{\rm for}\quad l<k
\end{array} \right. 
\eea
as our starting point in the study of the twist operator, and consequently of the mutual information as expressed in (\ref{mutual-corr}). We are interested in the $n\to 1$ limit of the above formula. This limit was considered in \cite{Casini:2021raa} and further examinated in \cite{Agon:2022efa}, where it was argued that
\bea \label{twist-correlator-3}
\langle \Omega |\tilde{\Sigma}^{(n)}_A {\cal O}^{l}(x_1) {\cal O}^{k}(x_2)|\Omega\rangle &\underset{n\to 1}{\approx} \tr\left\{\rho_A\,{ \cal O}(x_1){\cal O}_{A}\[x_2,i\frac{\tau_{kl}}{\pi}\] \right\}=\Big\langle{ \cal O}(x_1)  \,{\cal O}_{A}\[x_2,i\frac{\tau_{lk}}{\pi}\] \Big\rangle \,,
\eea
where  $\tau_{lk}\equiv \pi(l-k)/n$, and ${\cal O}_{A}\[x,s\]$ is the operator transformed by the modular flow of $A$, namely 
\bea\label{mod-evol}
{\cal O}_{A}\[x,s\]\equiv \rho_A^{-i s} {\cal O}(x)\rho_A^{i s}\,.
\eea
In (\ref{twist-correlator-3}) we assumed $l>k$. Notice that the identity part of the twist operator does not contribute to the correlator of two operators in different replicas as in our case. In the $n\to 1$ limit we are instructed to keep the $n$ dependence in the modular parameter, as the indices are bounded by $0\leq j\leq n$. The correlator that appear on the RHS of (\ref{twist-correlator-3}) is analytic in the strip with Im$\,s \in (0,1)$, and obeys the KMS periodicity between the boundaries of the strip \cite{Haag:1992hx}. This implies that the correlator is well defined on $0\leq \frac{\tau_{lk}}{\pi}\leq 1$. It follows from its definition that $1/n<\frac{\tau_{lk}}{\pi}<1-1/n$ provided we ordered the indices such that $l>k$. We will do so here.

There are few known cases in which the modular flow is known explicitly and whose action on local operators is geometrical. One such example is the modular flow for a spherical region in the ground state of a conformal field theory. For a sphere of radius $R$ centered at the origin, the modular flow acts via 
\bea\label{sphere-modular}
x^{\pm}_A[x,s]=R\frac{\(R+x^{\pm}\)-e^{\mp 2\pi s} \(R-x^{\pm}\) }{\(R+x^{\pm}\)+e^{\mp 2\pi s} \(R-x^{\pm}\)}\, ,
\eea
where $x^\pm=r\pm t$ are standard null coordinates \cite{ Brunetti:1992zf,Casini:2011kv,Hislop:1981uh}. This flow can be interpreted as a conformal transformation, which helps us derive the form in which the unitary acts on the operator.  We will consider scalar operators. The transformation is 
\bea\label{MF-CT}
{\cal O}_{A}[x,s]=U_A(s){\cal O}(x)U_A^{\dagger}(s)=\Omega_A^\Delta[x,s] {\cal O}(x_A[x,s]) \, ,
\eea 
where $U_A(s)$ is the unitary that implements the modular evolution, while $\Omega_A[x,s]$ is the conformal factor associated with the coordinate transformation, and we restrict in the following to primary operators. This can be derived from (\ref{sphere-modular}) and the relation $d x_A[s]^2=\Omega_A^2[x,s]d x^2$\,, or alternatively from the Jacobian of the transformation matrix. We find convenient to introduce the following simplified notation for such a Jacobian:
\bea
\left|\frac{\partial x_A(s)}{\partial x }\right|\equiv \left|\det\(\frac{\partial x^\alpha_A[x,s]}{\partial x^\beta }\)\right|^{\frac1d}=\Omega_A[x,s]\,.
\eea
Thus, the two-point function of interest has the form:
\bea\label{modular-corr}
\langle{ \cal O}(x_1){\cal O}_{A}\!\[x_2,s\]\rangle=\left|\frac{\partial x_{2;A}(s)}{\partial x_2 }\right|^\Delta \langle {\cal O}(x_1) {\cal O}(x_A[x_2,s])\rangle\,,
\eea
where 
 \bea\label{Jacobian}
 \left|\frac{\partial x_{2;A}(s)}{\partial x_2 }\right|=\frac{R^2 \sech^2(\pi s)}{(R-x_2^- \tanh(\pi s))(R+x_2^+ \tanh(\pi s))}\,.
 \eea
Our strategy in the exploration of the twist operator is to use (\ref{twist-correlator-3}) to deduce the form of the sector of the normalized twist operator $\tilde{\Sigma}_A^{(n)}$ that is responsible for that equality. We will solve this equation in two levels of complexity.

The simplest one is, what we call a {\sl local kernel} expansion, which entails the following ansatz
\bea\label{local-ansatz}
\tilde{\Sigma}_A^{(n)}= \sum_{l<k}\int_{D_A} {\cal G}^{lk}_A(\xi) {\cal O}^l(\xi){\cal O}^k(\xi)d^d\xi+\cdots
\eea 
for the sector of the twist operator, where $D_A$ denotes the domain of dependence associated to the entangling region $A$. This is the correct structure if we are interested in reproducing equation (\ref{twist-correlator-3}) in the particular case in which $x_1=x_2$. This is equivalent to reproducing all correlations of the twist with a single primary ${\cal O}^l(x){\cal O}^k(x)$ of the replicated theory, for any position $x$. In an OPE view of the twist operator, this will contain the contributions of the primary ${\cal O}^l{\cal O}^k$ and all its descendants, evaluated at the center of the sphere.
As we will see in section \ref{MI-Local}, this leads to a proof of the known conformal block expansion of the mutual information of disjoint spheres that have been induced from analogies with expansions of four point functions \cite{Long:2016vkg,Chen:2017hbk}. 

The other possibility is what we call the {\sl multi-local kernel} expansion. This entails the following ansatz 
\bea\label{multi-local-ansatz}
\tilde{\Sigma}_A^{(n)}= \sum_{l<k}\int_{D_A} \int_{D_A} {\cal G}^{lk}_A(\xi_1, \xi_2) {\cal O}^l(\xi_1){\cal O}^k(\xi_2)d^d\xi_1 d^d\xi_2+\cdots
\eea
for the associated sector of the twist operator. Using this ansatz one can reproduce (\ref{twist-correlator-3}) for arbitrary $x_1$ and $x_2$. As we will see in section \ref{MI-Multi-local}, this ansatz leads to a new expansion of the mutual information in terms of CFT primaries which effectively incorporates the contributions of an infinite set of replica primary operators. This set includes, not only ${\cal O}^l(\xi){\cal O}^k(\xi)$ and its descendants, but also all other primary operators that can be constructed by acting with an arbitrary number of derivatives on ${\cal O}^l(\xi)$ and ${\cal O}^k(\xi)$ independently, as well as their associated descendants. 
%Such resummation solves the problem of the proliferation of replica primaries. 

\subsection{N-copy sector}
The above ansatzes can be generalized to include the contributions of replica operators with support on an arbitrary number of replicas. This requires generalizing (\ref{twist-correlator}) for an arbitrary number of operator insertions. The appropriate generalization, assuming the indices $p_i$ are ordered, $n-1\geq p_1> \cdots > p_N\geq 0$, reads
\bea\label{twist-correlator-N}
\langle \Omega |\tilde{\Sigma}^{(n)}_A \prod_{j=1}^N {\cal O}^{p_j}_{\Delta_j}(x_j)| \Omega\rangle&=&
\tr\left\{\tilde{\rho}_A\,\prod_{j=1}^N\tilde{\rho}^{-\frac{(p_j-p_N)}n}_A\,{\cal O}_{\Delta_j}(x_j)\tilde{\rho}^{\frac{(p_j-p_N)}n}_A \right\}\,, \nonumber\\
&\underset{n\to 1}{\approx}&\expval{\prod_j \tilde{\mathcal{O}}_{\Delta_j,A}\left[x_j, -\frac{i \tau_{j N}}{\pi}\right]}\,,
\eea
where $\tau_{jN}=\pi (p_j-p_N)/n$. We can further use (\ref{MF-CT}) to simplify the right-hand side of (\ref{twist-correlator-N}) and re-write it as an $N-$point correlator times a series of conformal factors. 

 In this context, we can also consider both the local and multi-local kernel expansions for the twist operator, this is, the generalizations of (\ref{local-ansatz}) and (\ref{multi-local-ansatz}), respectively. For the  local kernel we write
\bea\label{local-ansatz-N}
\tilde\Sigma^{(n)}_A=\sum_{p_1 <\dots <p_N} \int_{D_A} d^{d} \xi\, \mathcal{G}_A^{p, \Delta}(\xi) \prod_{j=1}^N  \mathcal{O}^{p_j}_{\Delta_j}(\xi)\,,
\eea 
while for the  muti-local kernel we have
\bea\label{multi-local-ansatz-N}
\tilde\Sigma^{(n)}_A=\sum_{p_1 <\dots <p_N} \int_{D_A} d^{d} \xi_1\cdots \int_{D_A} d^d \xi_N\, \mathcal{G}_A^{p, \Delta}(\xi_1,\cdots, \xi_N) \prod_{j=1}^N  \mathcal{O}^{p_i}_{\Delta_i}(\xi_i)+\cdots\,,
\eea
where $p$ and $\Delta$ are compact notations for $\{p_1, \cdots, p_N\}$ and $\{\Delta_1, \cdots, \Delta_N\}$, respectively. In this work, we study in full detail the two-copy sector of the twist operator in the local and multi-local setups. We leave a thorough exploration of the N-copy sector contributions for future work.
%We present some preliminary results of the general structure of the N-copy sector contribution in section \ref{N-copy-sector} leaving a more detailed exploration for future work.

\section{Mutual information: Local kernel \label{MI-Local}}
In this section we study the ansatz (\ref{local-ansatz}) and the contribution to the mutual information that follows from it. Let us start by noticing that this ansatz can reproduce (\ref{twist-correlator-3}) only in the restricted case in which $x_1=x_2$. In this case, the function ${\cal G}^{lk}(\xi)$ (which we refer to as the local kernel) is defined through the equation
\bea\label{kernel-local-pq}
 \sum_{p<q}\int_{D_A} d^d\xi\,{\cal G}^{pq}_A(\xi) \langle {\cal O}^p(\xi){\cal O}^q(\xi){\cal O}^l(x){\cal O}^k(x)\rangle &=\langle{ \cal O}(x){\cal O}_{A}\[x,\frac{i \tau_{lk}}{\pi}\]\rangle\,,
\eea
where we take $x\in D_{\bar{A}}$, the spatial complement of $D_A$. 
Both sides can be computed in terms of two-point functions. Evaluating and simplifying each side of the above equality leads to \footnote{Notice that, going from \eqref{kernel-local-pq} to \eqref{eq:lk} the four-point function on the left hand side of \eqref{kernel-local-pq} reduces to a product of two-point functions since $p<q$, $l<k$ and the fact that the $n$ copies in CFT$^{\otimes n}$ are uncorrelated. We will use this fact in multiple places across the paper.} 
\bea\label{eq:lk}
\int_{D_A} d^d\xi\,{\cal G}^{kl}_A(\xi) \frac{1}{|x-\xi|^{4\Delta}}= \frac{1}{4^\Delta \sinh^{2\Delta}\(i \tau_{lk}\)}\frac{|x_A^+ -x_A^-|^{2\Delta}}{|x -x_A^-|^{2\Delta}|x -x_A^+|^{2\Delta}}\,,
\eea
where we have written the right-hand side of the above formula in terms of correlators that involve the future and past tips of the causal diamond associated to $A$, which we denoted respectively by $x_A^+=R_A\hat{t}$ and $x_A^-=-R_A\hat{t}$, with $\hat{t}$ the unit vector that defines the time direction. In appendix \ref{OPE-blocks}, by studying an integral representation of a scalar conformal block we find a  solution to (\ref{eq:lk}) in (\ref{rel-corr-1}), which is given by 
\bea\label{local-kernel}
\,{\cal G}^{kl}_A(\xi)=\frac{\tilde{c}_{2\Delta}}{4^\Delta \sinh^{2\Delta}\(i \tau_{lk}\)}\(\frac{|x_A^+ -x_A^-|}{|\xi -x_A^-||\xi -x_A^+|}\)^{2\Delta-d}\,.
\eea
The normalization coefficient $\tilde{c}_{2\Delta}$ is given by (\ref{coeff-calpha}), which we reproduce here  for convenience
\bea
\tilde{c}_{2\Delta}=\frac{4^{\Delta}\Gamma\(\Delta+\frac{1}2\)\Gamma\(2\Delta+1-\frac{d}2\)}{\pi^{\frac{d-1}2}\Gamma\(\Delta\) \[\Gamma\(\Delta+1-\frac{d}{2}\)\]^2}\,.
\eea
Although, (\ref{local-kernel}) may not be unique, adding a homogeneous term ---a term that vanishes under the integral (\ref{eq:lk})--- does not change its contribution to the mutual information. This can be seen from the fact that the final formula for the mutual information contains precisely (\ref{eq:lk}). Therefore, the solution (\ref{local-kernel}) suffices for our purposes. Plugging (\ref{local-kernel}) into (\ref{local-ansatz}) leads to 
\bea
\tilde{\Sigma}_A^{(n)}\approx \sum_{k<l}\frac{1}{4^\Delta \sinh^{2\Delta}\(i\tau_{lk}\)}\[\tilde{c}_{2\Delta}\int_{D_A} \(\frac{|x_A^+ -x_A^-|}{|\xi -x_A^-||\xi -x_A^+|}\)^{2\Delta-d} {\cal O}^l(\xi){\cal O}^k(\xi)d^d\xi\]\,,
\eea
where the term in brackets can be identified with the OPE block associated with the replica primary operator ${\cal O}^l(\xi){\cal O}^k(\xi)$, see equation (\ref{OPE-int-rep-6}).  Using (\ref{mutual-corr}) we obtain a formula for its contribution to the mutual information 
\bea\label{Mutual-CB-scalar}
I^{\ell}_\Delta(A:B)=\lim_{n\to 1}\frac{1}{n-1}\sum_{k<l}\frac{1}{2^{4\Delta} \sin^{4\Delta}\( \tau_{lk}\)}G_{2\Delta}(u,v)=\frac{1}{2^{4\Delta}}\frac{\sqrt{\pi} \Gamma[2\Delta+1]}{4\Gamma[2\Delta+\frac{3}{2}]}G_{2\Delta}(u,v)
\eea
where $G_{2\Delta}(u,v)$ is the conformal block associated to a scalar operator of dimension $2\Delta$ and $u, v$ are the standard conformal ratios
\bea\label{conf-parameters-1}
u\equiv \frac{|x_1-x_2|^2|x_3-x_4|^2}{|x_1-x_3|^2|x_2-x_4|^2}\,,\quad{\rm and }\quad  v\equiv \frac{|x_1-x_4|^2|x_2-x_3|^2}{|x_1-x_3|^2|x_2-x_4|^2}\,,
\eea
where the points $x_i$ are taken to be the tips of the causal cones associated to the two spheres. This is  
$x_1=x_A^-$, $x_2=x_A^+$  and $x_3=x_B^-$, $ x_4=x_B^+$. In the last equality in (\ref{Mutual-CB-scalar}), we used the result of the following sum
\bea\label{scalar-sum}
\lim_{n\to 1}\frac{1}{n-1}\sum_{k<l}\frac{1}{\sin^{4\Delta}\(\tau_{lk}\)}=\lim_{n\to 1}\frac{n}{n-1} \frac{1}{2}\sum_{l=1}^{n-1}\frac{1}{\sin^{4\Delta}\( \frac{\pi l}{n}\)}=\frac{\sqrt{\pi} \Gamma[2\Delta+1]}{4\Gamma[2\Delta+\frac{3}{2}]}\,.
\eea
The subindex $\Delta$ in (\ref{Mutual-CB-scalar}) indicates that we only consider the contribution to the mutual information related to the primary operator of dimension $\Delta$.

\subsection{Conformal block expansion}

The procedure that led to (\ref{Mutual-CB-scalar}) can be straightforwardly generalized to include arbitrary replica operators, that is, replica operators with support on an arbitrary number of replica copies. This includes operators of the form (\ref{local-ansatz-N}). Consequently, the twist operator can be expanded in terms of OPE blocks of replica primaries, as originally proposed in \cite{Long:2016vkg}. This proposal is based on an analogy: because an operator on a $d-2$-dimensional spatial sphere is determined by the two tips of the causal diamond, it was proposed that the operator can be expanded in OPE blocks as the product of two field operators at the tips. This expansion, however, does not satisfy some positivity relations coming from unitarity that holds for the product of two fields. Here we showed that this proposal is indeed correct, not because the twist operator can be assimilated to two fields but because, as any localized operator, it has to have an expansion in terms of primaries. By causality, this expansion is supported in $D_A$, and restrictions imposed by conformal invariance give the coefficients as the OPE blocks. Therefore, the mutual information of disjoint spheres in a conformal field theory can be expanded in terms of conformal blocks:
\bea\label{MI-CB}
I(A:B)=\sum_{\Delta,J}b_{\Delta,J} G^d_{\Delta,J}(u,v)\, ,
\eea
where the coefficients $b_{\Delta,J}$ can be determined from similar sums as the one in (\ref{scalar-sum}) or via an integral over the modular parameter following the methods of \cite{Agon:2015ftl, Casini:2021raa}. Indeed, the above expansion for the mutual information is well known in the literature; it was proposed in \cite{Chen:2017hbk}, and studied in various concrete examples in \cite{Chen:2017hbk, Chen:2016mya}.

Unfortunately, this formula has some disadvantages. First, the classification of replica operators that follow from a given seed CFT, and for an arbitrary number of replicas, is a formidable problem. Second, even if one solves this problem, the computation of their associated coefficients
 $b_{\Delta,J}$ is also technically challenging. From our current perspective, the expansion in terms of OPE blocks is a consequence of imposing (\ref{twist-correlator-3}) constrained to the diagonal part $x_1=x_2$. A much more powerful expansion is expected if one instead imposes the full non-diagonal equation (\ref{twist-correlator-3}), which is possible if one starts from the ansatz (\ref{multi-local-ansatz}). Indeed, in the next section, we will show how this latter ansatz solves both problems in the particular case of the two-copy sector of the replica theory.

%In this section we would like to present a new perspective on the origin of the above formula which helps to clarify its meaning and suggest a generalization which solves one of the practical difficulties that this formula entails, namely the proliferation of replica primaries. 

\section{Mutual information: Multi-local kernel \label{MI-Multi-local}}

In this section we study the bi-local ansatz (\ref{twist-correlator-3}) and the contribution to the mutual information that follows from it. This ansatz allow us to solve equation (\ref{twist-correlator-3}) with arbitrary $x_1$ and $x_2$ both in $D_{\bar{A}}$, the spatial complement of $D_{A}$.
The equation of interest is
\bea\label{bi-local-eq-1}
 \sum_{p<q}\int_{D_A} d^d\xi_1\int_{D_A}d^d\xi_2\,{\cal G}^{pq}_A(\xi_2,\xi_1) \langle {\cal O}^p(\xi_2){\cal O}^q(\xi_1){\cal O}^l(x_1){\cal O}^k(x_2)\rangle &=\langle{ \cal O}(x_1){\cal O}_{A}\[x_2,\frac{i \tau_{lk}}{\pi}\]\rangle\,,
\eea
where once again we look for a function ${\cal G}^{pq}_A(\xi_1,\xi_2) $ for which the above equality holds. Simplifying each side of the equation we get 
\bea\label{bi-local-eq-2}
\int_{D_A} d^d\xi_1\int_{D_A}d^d\xi_2\,{\cal G}^{kl}_A(\xi_2,\xi_1) \frac{1}{|\xi_1-x_1|^{2\Delta}}\frac{1}{|\xi_2-x_2|^{2\Delta}}=\left|\frac{\partial x_{2;A}(i\frac{\tau_{lk}}\pi)}{\partial x_2 }\right|^\Delta\frac{1}{\left|x_1-x_{2;A}\(\frac{i \tau_{lk}}{\pi}\)\right|^{2\Delta}}\,,
\eea
where the Jacobian factor above is given by  (\ref{Jacobian})\,. Formally, one can imagine obtaining the kernel ${\cal G}^{pq}_A(\xi_1,\xi_2)$ by using an inverse operator for the ordinary CFT two-point function but with insertion points at $D_{A}$ and $D_{\bar{A}}$ respectively.  Thus, we conjecture that the inverse operator with properties 
\bea\label{inverse-prop-1}
&&\int_{D_A} d^d \xi_1 \frac{G^{-1}_{A;\Delta} (\xi_1,w_1)}{| \xi_1-\chi_1|^{2\Delta}}=\delta^{d}(w_1-\chi_1)\quad{\rm where}\quad \{w_1, \chi_1\}\in D_{\bar{A}}, \quad {\rm and} \\ \label{inverse-prop-2}
&&\int_{D_{\bar{A}}} d^d w_1 \frac{G^{-1}_{A;\Delta} (\xi_1,w_1)}{| w_1-\zeta_1|^{2\Delta}}=\delta^{d}(\xi_1-\zeta_1)\quad{\rm where}\quad \{\xi_1, \zeta_1\}\in D_A
\eea
exist  for any spherical region $A$. The subindex $A$ in the inverse operator reminds us that such operator depends on the entangling region. Applying such inverse operators to equation (\ref{bi-local-eq-2}) leads to an explicit expression for ${\cal G}^{kl}_A(\xi_2,\xi_1)$
\bea\label{bi-local-kernel}
{\mathcal G}^{kl}_A(\xi_2, \xi_1)=\int_{D_{\bar{A}}} d^d x_1 \int_{D_{\bar{A}}}  d^d x_2 \, G^{-1}_{A;\Delta} (\xi_1,x_1)\,G^{-1}_{A;\Delta} (\xi_2,x_2) 
\left|\frac{\partial x_{2;A}\(i\frac{\tau_{lk}}{\pi}\)}{\partial x_2}\right|^{\Delta}\frac{1}{\left|x_1-x_{2;A}\(\frac{i \tau_{lk}}{\pi}\)\right|^{2\Delta}}\,.\nonumber\\
\eea
As in the local kernel solution (\ref{eq:lk}), one can add a homogeneous term to the above expression but this does not change its contribution to the mutual information. The above equation could be further simplified by evaluating the integral over $x_1$ from property (\ref{inverse-prop-2}). However, such evaluation involves a delta function evaluated on a complex argument. To avoid dealing with such objects we keep the form of the above formal expression and study the mutual information assuming knowledge of ${\mathcal G}^{kl}_A(\xi_1, \xi_2)$. From (\ref{multi-local-ansatz}) and (\ref{mutual-corr}) we have 
\bea\label{mutual-lorentzian-1}
I^{n\ell}_\Delta(A:B)=\lim_{n\to 1}\frac{1}{n-1}\sum_{l<k}&& \int_{D_A} d^d \,\xi_1 \int_{D_A} d^d \xi_2\, \int_{D_B} d^d \chi_1 \int_{D_B} d^d \chi_2\,\frac{{\mathcal G}^{lk}_A(\xi_1, \xi_2)\,{\mathcal G}^{lk}_B(\chi_1, \chi_2)}{\left| \chi_1-\xi_1+L\right|^{2\Delta}\left| \chi_2-\xi_2+L\right|^{2\Delta}}  \nonumber\\
\eea
where for simplicity we used coordinates $\{\xi_1,\xi_2\}$ and $\{\chi_1,\chi_2\}$ adapted to the spheres $A$ and $B$, respectively. Thus, we explicitly put the separation between the spheres $L$ in the correlators. The kernels ${\mathcal G}^{lk}_{A/B}$ as expressed in (\ref{bi-local-kernel}), are given in terms of integrals over a modular evolved correlator (\ref{modular-corr}). Since such correlator is analytic on the strip defined by $0<\Re(\tau_{lk})<\pi$ (as discussed around (\ref{mod-evol})), and assuming such analyticity is not lost after the multiple integrations involved in (\ref{mutual-lorentzian-1}) and (\ref{bi-local-kernel}), we might conclude that the function we are summing over $\{l,k\}$ in (\ref{mutual-lorentzian-1}) is analytic. Under this assumption, we can follow the methods of \cite{Agon:2015ftl,Casini:2021raa} and continue $\tau_{lk}$ into a complex variable. The sum over $\{l,k\}$ can be turned into an integral over the modular parameter $s$ which is obtained after the transformation:
\bea\label{analytic-cont}
i\frac{\tau_{lk}}{\pi}\to \frac{i}{2}+s \quad \Rightarrow \quad \lim_{n\to 1}\frac{1}{n-1}\sum_{l<k} \to \frac{1}4 \int_{-\infty}^{\infty} \frac{\pi \, ds}{\cosh^2(\pi s)}\,.
\eea
 Under this transformation the point $x_{2;A}\(\frac{i \tau_{lk}}{\pi}\) \to x_{2;A}\(\frac{i}2+s\)$. As a consequence, $x_{2;A}(s)$ which lies in $D_{\bar{A}}$ for all real $s$, gets mapped to its causal complement $D_A$ under the $i/2 $ shift. Thus, we introduce the following notation 
\bea
x_{2;A}^J(s)\equiv x_{2;A}\(\frac{i}2+s\)\quad{\rm where}\quad  x_{2;A}^J(s)= x_A\[x_{2;A}^J,s\]
\eea
and by $x_{2;A}^J$ we mean the point that is obtained under the action of the Tomita operator $J$ which implements the modular evolution by $i/2$ for region $D_A$ \cite{Haag:1992hx}.  The precise map is given by\footnote{This map should be used with care since it is only well defined in a subregion of the causal diamond in the Minkowski space-time. Fortunately, \eqref{J-action} is local and well defined everywhere in the cylinder compactification of Minkowski space-time \cite{Segal1971, Segal1976, Luscher:1974ez}. Therefore, in explicit computations one should use the above map in the cylinder and translate the results back to Minkowski space only at the end of the computations. See appendix \ref{Tomita-J} for a discussion of the subtleties related to the map. } 
\begin{equation}\label{J-action}
  x^{J,0}_{2,A}= -\frac{R_A^2 x_2^0}{|x_2|^2}\,, \quad {\rm and }\quad  x^{J,i}_{2,A}=  \frac{R_A^2 x_2^i}{|x_2|^2}\,.
\end{equation}
The transformation (\ref{analytic-cont}) makes $x_{2;A}\(\frac{i}2+s\)$ real, which simplifies the form of the functions ${\mathcal G}^{lk}_{A/B}(\xi_1,\xi_2)$, now called ${\mathcal G}_{A/B}(\xi_1,\xi_2;s)$. Namely, we can carry out the integral over $x_1$ in (\ref{bi-local-kernel}) using the inverse operator, which results in
\bea\label{bi-local-kernel-2}
{\mathcal G}_A(\xi_2, \xi_1;s)=\int_{D_{\bar{A}}} d^d x_2\,G^{-1}_{A;\Delta} (\xi_2,x_2) 
\left|\frac{\partial x^J_{2;A}\(s\)}{\partial x_2}\right|^{\Delta}\delta^d\({\xi_1-x^J_{2;A}\(s\)}\) \,.
\eea
The remaining integral can be carried out explicitly, using the delta function by changing the integration variable from $x_2$ to $x_{2;A}^J(s)$. This is possible for fixed $s$ but entails a change in the integration region from $D_{\bar{A}}$ to $D_{A}$. The explicit change of variables is given by $x_2=x_A[x_{2;A}^J(s),-\frac{i}{2}-s]$, and the result of the integral is
\bea\label{bi-local-kernel-3}
{\mathcal G}_A(\xi_2, \xi_1;s)= 
\left|\frac{\partial \xi^J_{1;A}\(-s\)}{\partial \xi_1}\right|^{d-\Delta}G^{-1}_{A;\Delta} (\xi_2,\xi_1^{J}(-s))\,.
\eea
It is remarkable that we can write down an explicit formula for the bi-local kernel of the twist operator. 

 After the analytic continuation (\ref{analytic-cont}), the expression for the mutual information becomes
\bea\label{mutual-bi-local-2}
I^{n\ell}_\Delta(A:B)=\frac{1}{4}\int_{-\infty}^\infty \frac{\pi \,ds}{\cosh^2\pi s} && \int_{D_A} d^d \,\xi_1 \int_{D_A} d^d \xi_2\, \int_{D_B} d^d \chi_1 \int_{D_B} d^d \chi_2\,\frac{{\mathcal G}_A(\xi_1, \xi_2;s)\,{\mathcal G}_B(\chi_1, \chi_2;s)}{\left| \chi_1-\xi_1+L\right|^{2\Delta}\left| \chi_2-\xi_2+L\right|^{2\Delta}}\,.  \nonumber\\
\eea
To evaluate the above integral we find it more convenient to use the expression (\ref{bi-local-kernel-2}) for the kernel ${\mathcal G}_A(\chi_1, \chi_2;s)$ and the following equivalent expression for ${\mathcal G}_B(\chi_1, \chi_2;s)$,
\bea\label{bi-local-kernel-3}
{\mathcal G}_B(\chi_2, \chi_1;s)=\int_{D_{\bar{B}}} d^d x_1\,G^{-1}_{B;\Delta} (\chi_1,x_1) 
\left|\frac{\partial x^J_{1;B}\(-s\)}{\partial x_1}\right|^{\Delta}\delta^d\({\chi_2-x^J_{1;B}\(-s\)}\)\,. 
\eea 
This later formula follows from the identity $\langle{ \cal O}(x_1){\cal O}_{B}\[x_2,s\]\rangle=\langle{ \cal O}_B\[x_1,-s\]{\cal O}(x_2)\rangle$ applied to (\ref{bi-local-kernel}), which allow us to carry out the integral over $x_2$ instead of the one over $x_1$.

\subsection{Long-distance limit}

Before studying (\ref{mutual-bi-local-2}) in detail, it is convenient to analyze the long-distance limit of that formula. This can be achieved by taking the $R_A, R_B\to 0$ limit. In such case, one can pull out the correlators from the integrands, which turns (\ref{mutual-bi-local-2}) into 
\bea\label{long-dist-mutual}
I^{n\ell}_\Delta(A:B)=\frac{1}{L^{4\Delta}}\int_{-\infty}^\infty \frac{\pi \,ds}{4\,\cosh^2\pi s}c^{A}_{\Delta}(s)c^{B}_{\Delta}(s)\,=c_\Delta\frac{R_A^{2\Delta}R_B^{2\Delta}}{L^{4\Delta}}\,, 
\eea 
where 
\bea\label{cA-normal}
c^{A}_{\Delta}(s)=\int_{D_{A}} \!\!\! d^d\xi_1 \int_{D_{A}}\!\!\!d^d\xi_2\, {\mathcal G}_A(\xi_2, \xi_1;s)\,,
\eea
and similarly for $c^{B}_{\Delta}(s)$.

Equation (\ref{long-dist-mutual}) tells us that at very long separation distances the mutual information decays as the square of the correlator of the lowest scaling dimension operator in the theory. This fact is well-known \cite{Headrick:2010zt,Cardy:2013nua}  as well as the coefficient in front \cite{Calabrese:2010he,Agon:2015ftl}. In our framework, we can calculate that coefficient by direct integration of (\ref{cA-normal}) and (\ref{long-dist-mutual}). First, we can integrate over $\xi_1$ using the delta function in (\ref{bi-local-kernel-2}) since $x^J_{2;A}(s)\in D_A$. To be able to integrate over $x_2$, notice that the Jacobian factor is proportional to a spacelike correlator, this is
 \bea\label{Jacobian-corr}
 \left|\frac{\partial x^J_{A}(-s)}{\partial x}\right|^\Delta=\frac{R_A^{2\Delta}}{\cosh^{2 \Delta}\pi s}\frac{1}{|R_A(s)-x|^{2\Delta}}\,,
 \eea
where $R_A(s)=R_A \tanh \pi s\,  \hat{t}$, which follows from (\ref{Jacobian}). Therefore, using property (\ref{inverse-prop-2}) the integral over $x_2$ leads to a delta function which allows us to integrate over $\xi_2$. Thus, the coefficient $c^{A}_{\Delta}(s)$ is simply given by 
\bea
c^{A}_{\Delta}(s)=\frac{R_A^{2\Delta}}{\cosh^{2\Delta}\pi s}\,,
\eea
and the coefficient in the long-distance mutual information (\ref{long-dist-mutual}), is 
\bea
c_\Delta= \int_{-\infty}^\infty \frac{\pi \,ds}{4\,\cosh^{4\Delta+2}\pi s}=\frac{\sqrt{\pi}\,\Gamma\[2\Delta+1\]}{4\,\Gamma\[2\Delta+\frac32\]}\,,
\eea
which matches the known results \cite{Calabrese:2010he,Agon:2015ftl}.

\subsection{General calculation}
We now proceed to evaluate the expression (\ref{mutual-bi-local-2}) in full generality. The delta functions in equations (\ref{bi-local-kernel-2}) and (\ref{bi-local-kernel-3}), once in (\ref{mutual-bi-local-2}) allow us to evaluate the integrals over $\xi_1$ and $\chi_2$. Similarly, the inverse correlators in equations (\ref{bi-local-kernel-2}) and (\ref{bi-local-kernel-3}) allow us to integrate over $\xi_2$ and $\chi_1$, respectively. All in all, we get 
\bea\label{mutual-bi-local-3}
I^{n\ell}_\Delta(A:B)=\frac{1}{4}\int_{-\infty}^\infty \frac{\pi \,ds}{\cosh^2\pi s} && \int_{D_{\bar{A}}} d^d x_2 \int_{D_{\bar{B}}} d^d x_1\,\left|\frac{\partial x^J_{2;A}\(s\)}{\partial x_2}\right|^{\Delta}\left|\frac{\partial x^J_{1;B}\(-s\)}{\partial x_1}\right|^{\Delta} \nonumber \\
&& \qquad\qquad \times \delta^d\(x_1-x^J_{2;A}(s)+L\) \delta^d\(x_2-x^J_{1;B}(-s)-L\)\,.  
\eea
The integral over $x_1$ can be done trivially, leading to 
\bea\label{mutual-bi-local-4}
I^{n\ell}_\Delta(A:B)=\frac{1}{4}\int_{-\infty}^\infty \frac{\pi \,ds}{\cosh^2\pi s}  \int_{D_{\bar{A}}}\!\!  d^d x_2\,
\left|\frac{\partial x^J_{2;A}(s)}{\partial x_2}\right|^{\Delta} \!\! && \!\!
 \left|\frac{\partial x^J_{1;B}(-s)}{\partial x_1}\right|^{\Delta}_{x_1=x^J_{2;A}(s)-L} \!\!\! \!\!\!\!\! \!\!\!\!\! \!\!\!\!\!\!\!\!\!\!\!\!
\delta^d\( x_2-L-x^J_B\[x^J_{2;A}(s)-L,-s\]\)  \nonumber\\
&&\!\!\!\!\!\!\!\!\!\!\times \Theta\(x_{2;A}^J(s)-L\in D_{\bar{B}} \)\Theta\(x_2-L\in D_{B} \)\,. 
\eea
Finally, the above integral can be done by using the last delta function and solving for its zeros, including the appropriate Jacobian of the required coordinate transformation, this is 
 \bea\label{mutual-bi-local-5}
I^{n\ell}_\Delta(A:B)=\frac{1}{4}\int_{-\infty}^\infty \frac{\pi \,ds}{\cosh^2\pi s}  \!\! && \!\sum_{i}
\left|\frac{\partial x^J_{2;A}(s)}{\partial x_2}\right|_{x_2=w_i^*}^{\Delta} 
 \left|\frac{\partial x^J_{1;B}(-s)}{\partial x_1}\right|^{\Delta}_{x_1=x^J_A[w_i^*,s]-L} \nonumber\\
&& \times 
\frac{\Theta\(x_{2;A}^J(s)-L\in D_{\bar{B}} \)\Theta\(x_2-L\in D_{B} \) \Big|_{x_2=w_i^*}}{\left|\det \(\frac{\partial g^\alpha(x_2)}{\partial x_2^\beta}\)\right|_{x_2=w_i^*}}\,, 
\eea
where $w_i^*$ are the zeros of the function $g(x_2)$ defined as 
\bea\label{zeros-g}
 g(x_2)=x_2-L-x^J_B\[x^J_{A}[x_2,s]-L,-s\]\,.
 \eea
Such zeros come in four branches, these are
\begin{align}\label{wsols}
w_{1,\pm}^*(s)&=\left\{-R\tanh\pi s\pm \frac{L}2 \sqrt{1- \frac{4 R^2 }{L^2 \cosh^2 \pi s}} , \frac{L}{2}\right\}\,,\\
w_{2,\pm}^*(s)&=\left\{-R \tanh \pi s, \frac{L}{2} \pm \frac{L}{2} \sqrt{1- \frac{4 R^2 }{L^2 \cosh^2 \pi s}} \right\}\,.
\end{align}
In the above expression, the first term is the time component, and the second is the spatial component along the direction of $\vec{L}$. 

\begin{figure}
\begin{center}
\includegraphics[scale=0.45]{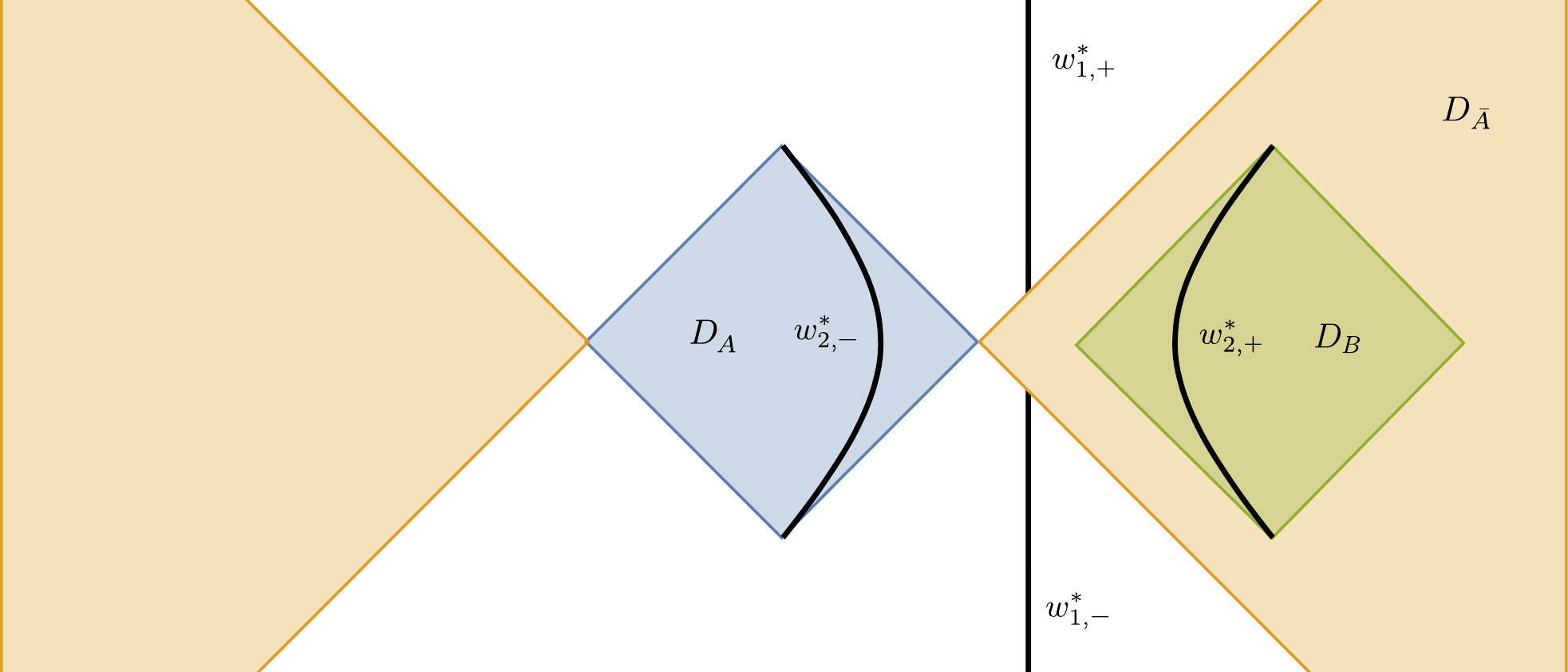}
\end{center}
\caption{Zeroes of $g(w)$ in Lorentzian signature. The four disconnected curves represent the four branches of solutions described in (\ref{wsols}).} 
\label{fig:plot_lorentzian}
\end{figure}

\begin{figure}
\begin{center}
\includegraphics[scale=0.45]{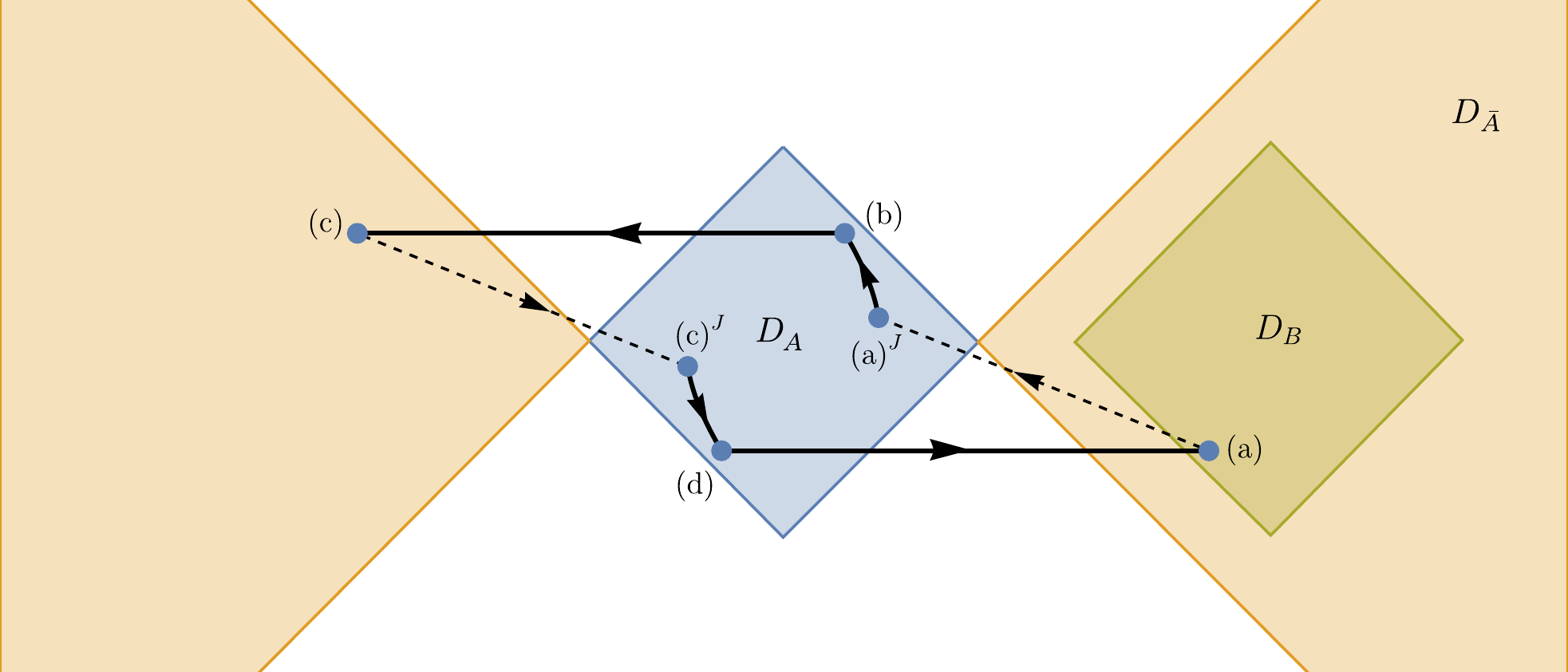}
\end{center}
\caption{Graphic representation of (\ref{zeros-g}) via a sequence of modular transformations (flow and $J$ conjugation ) and translations that close a loop in real space. Point (a) represents the solution $w_{2,+}^*(s)$ for some value of $s$ and (a)$^J$ is the point obtained after conjugation by $J$. Point (b) is the image under a modular flow generated by $D_A$ and amount $s$, $x^J_A[w_{2,+}^*(s),s]$. Point (c) is obtained from (b) by a $-L$ translation, (c)$^J$ is its $J$-conjugate point while point (d) is obtained from (c)$^J$ by the modular flow of $B$ an amount $-s$, $x^J_{B}[x^J_A[w_{2,+}^*(s),s]-L,-s]$. Finally, a further translation of (d) by $L$ takes (d) into the initial point $w_{2,+}^*(s)$ by virtue of (\ref{zeros-g}). The above geometry suggests that the mutual information is built from the modular commutativity between regions $A$ and $B$.} 
\label{fig:mod-commutativity}
\end{figure}
Figure \ref{fig:plot_lorentzian} provides a graphic representation of the four branches of solutions (\ref{wsols}) within the physical configuration, namely, the space where the causal diamond $D_A$ and its spatial complement lives. We also represent the actual location of the causal diamond $D_B$\footnote{The reader should remember that all of our formulas use coordinates adapted to $D_B$. Therefore $D_B$ and its modular evolution are computed for a centered sphere}. This figure shows that only the solution $w^*_{2,+}$ contributes to the mutual information due to the theta functions. It is straightforward to check that both theta functions leave this solution untouched. It is interesting to notice that (\ref{mutual-bi-local-5}) is a sum of contributions that are determined from a sort of modular commutativity condition (\ref{zeros-g}) for each value of the modular parameter $s$. Such a condition couples the modular flows of both regions $A$ and $B$ and includes the separation distance $L$. We illustrate this geometry in figure \ref{fig:mod-commutativity}. 

To evaluate $I^{n \ell}_\Delta(A:B)$, we need to compute the determinants in (\ref{mutual-bi-local-5}),
on $w^*_{2,+}$. For that purpose, it is convenient to re-write the expression of interest in terms of the transformed point
\bea\label{sol-tw}
\tilde{w}^*(s):=x_A^J[w^*_{2,+},s]=\left\{-R \tanh \pi s, \frac{L}{2} - \frac{L}{2} \sqrt{1- \frac{4 R^2 }{L^2 \cosh^2 \pi s}} \right\}=w^*_{2,-}(s)\,.
\eea
Equation (\ref{zeros-g}) becomes
\bea
 f(\tilde{w})=x_A^J[\tilde{w},-s]-L-x^J_B\[\tilde{w}-L,-s\]\,,
\eea
where (\ref{sol-tw}) is the solution to $f(\tilde{w})=0$ we are interested in. Thus, (\ref{mutual-bi-local-5}) becomes
\bea\label{mutual-bi-local-6}
I^{n\ell}_\Delta(A:B)=\frac{1}{4}\int_{-\infty}^\infty \frac{\pi \,ds}{\cosh^2\pi s} \left|\frac{\partial \tilde{w}^J_A(-s)}{\partial \tilde{w}}\right|_{\tilde{w}=\tilde{w}^*}^{d-\Delta} 
 \left|\frac{\partial x^J_{1,B}(-s)}{\partial x_1}\right|^{\Delta}_{x_1=\tilde{w}^*-L} \,
\frac{1}{\left|\det \(\frac{\partial f^\alpha(\tilde{w})}{\partial \tilde{w}^\beta}\)\right|_{\tilde{w}=\tilde{w}^*}}\,. 
\eea
The Jacobian factors above can be evaluated explicitly using  (\ref{Jacobian}) and (\ref{sol-tw}), which results in
\bea
&&\left|\frac{\partial \tilde{w}^J_A(-s)}{\partial \tilde{w}}\right|_{\tilde{w}=\tilde{w}^*}^{d-\Delta}=\frac{\(\cosh\pi s+\sqrt{\cosh^2\pi s-z}\)^{2(d-\Delta)}}{z^{d-\Delta}}\,, \\
&& \left|\frac{\partial x^J_{1,B}(-s)}{\partial x_1}\right|^{\Delta}_{x_1=\tilde{w}^*-L}\!\!\! =\frac{z^{\Delta}} {\(\cosh\pi s+\sqrt{\cosh^2\pi s-z}\)^{2 \Delta}}\,,
\eea
where $z=4R^2/L^2$, is the conformal ratio. The evaluation of the determinant requires extra work, but the final answer is simply given by\footnote{The interested reader can find the details of this calculation in Appendix \ref{det-lorentzian}.}
\bea
\left|\det \(\frac{\partial f^\alpha(\tilde{w})}{\partial \tilde{w}^\beta}\)\right|_{\tilde{w}=\tilde{w}^*}=\frac{4^d \cosh^d\pi s \(\cosh^2\pi s-z\)^{\frac{d}{2}}}{z^d}\,.
\eea
 Putting everything together we get the contribution of the mutual information 
\bea
I^{n\ell}_\Delta(A:B)&=&\frac{z^{2\Delta}}{4^{d+1}}\,\int_{0}^\infty \frac{2\pi\, ds}{\cosh^{d+2}\pi s} \frac{\left(  \cosh \pi s + \sqrt{\cosh^2\pi s-z}\right)^{2(d-2\Delta)}}
{ \(\cosh^2\pi s-z \)^{\frac{d}2}  } \,,
\eea
where we used the fact that the integrand is even in $s$. We can rewrite the above contribution in many ways. One particular form that we found useful is obtained by changing the integration variable $w \to1/\cosh^2\pi s $ resulting in
\bea\label{int-rep-2}
\boxed{I^{n\ell}_\Delta(A:B)=\frac{z^{2\Delta}}{4^{d+1}}\,\int_{0}^1 \frac{ w^{2\Delta}}{\sqrt{1-w}} \frac{\left(  1 + \sqrt{1-w\,z}\right)^{2(d-2\Delta)}}
{ \(1-w\,z \)^{\frac{d}2}  }\,dw\,.}
\eea
Expanding the binomial inside the integral, using the fact that the resulting series is convergent, one can rewrite the above formula as an infinite sum of Gaussian hypergeometric functions\footnote{A useful integral representation of the Hypergeometric function is
\bea\label{int-rep-2F1}
\,_2F_1\(a,b,c;z\)=\frac{\Gamma\( c\)}{\Gamma\(b\)\Gamma\(c-b\)}\int_0^1 w^{b-1} (1-w)^{c-b-1}(1-z w)^{-a}dw\,,
\eea
which is valid for $\Re\{c\}>\Re{b}>0$. Comparing with our formula of interest, we find $a=(d-k)/2$, $b=2\Delta+1$ and $c=2\Delta+3/2$.
} 
\bea\label{I-hypers}
\boxed{I^{n\ell}_\Delta(A:B)
=\frac{z^{2\Delta}}{4^{d+1}}\frac{ \sqrt{\pi}\,\Gamma\(2\Delta+1\)}{\Gamma\(2\Delta+\frac32\)}\sum_{k=0}^{\infty}{{2(d-2\Delta)} \choose k}\,\,_2F_1\(\frac{d-k}2,2\Delta+1,2\Delta+\frac32;z\)\,,}
\eea
which is valid for $\Delta>-\frac{1}{2}$. In particular, the sum over the binomial coefficients truncate for $2(d-2\Delta)\in \mathbb{Z}^+$ and it is well defined for $2(d-2\Delta)\notin \mathbb{Z}^{-}$ which means $\Delta\notin \frac{d}2+\frac14\mathbb{Z}^+$. As we will see in the following subsections both (\ref{int-rep-2}) and (\ref{I-hypers}) are useful expressions depending on the particular application. These equations represent the main result of the paper. In an OPE analysis of the twist operators they give the contribution of all operators that can be constructed with the field ${\cal O}_\Delta$ in two copies. This includes all primaries that can be formed by combining these two operators and their descendants, and all descendants of these primaries. In terms of conformal blocks, these expressions are a resumation of an infinite number of conformal blocks corresponding to the different two copy primaries generated by ${\cal O}_\Delta$.  

\subsection{Expansion at long distances \label{App-A}} 
In this subsection, we will study the long-distance expansion of  (\ref{I-hypers}) and compare it with the analogous expansion that follows from the more familiar conformal blocks decomposition (\ref{MI-CB}). For that purpose, we use the series expansion of the hypergeometric functions for small $z$. Thus, we write  
\bea\label{I-small-x-series}
I^{n\ell}_\Delta(A:B)&=&\(\frac{z}{4}\)^{2\Delta}\sum_{n=0}^{\infty} \(\frac{z}{4}\)^n\frac{ \sqrt{\pi}\,\Gamma\(2\Delta+1+n\)}{4\, \Gamma\(2\Delta+\frac32+n\)}K_\Delta(d,n)
\eea
where 
\bea
K_\Delta(d,n)\equiv\frac{4^{2\Delta+n-d}}{n!}\sum_{k=0}^{\infty}{{2(d-2\Delta)} \choose k}\,\frac{\Gamma\(\frac{d-k}{2}+n\)}{\Gamma\(\frac{d-k}{2}\)}\,.
\eea
The convergence of $I^{n \ell}_\Delta(A:B)$ for fixed $z$ allows us to exchange the order of the sums and write down the above compact expression. 
The sum in the definition of $K_\Delta(d,n)$ can be written down in terms of hypergeometric $_3F_2$ functions evaluated at one as
\bea
K_\Delta(d,n)
 &=&\frac{4^n\Gamma\( \frac{d}2 +n\)}{2n!\,\Gamma\(\frac{d}2\)}\,_3F_2\[\begin{array}{c}
-n, \,2\Delta-d, \,2\Delta-d+\frac12 \\
 1-2d +4 \Delta,\,1-\frac{d}2-n
 \end{array}
 ; 1  \] \nonumber\\
 &&+\frac{4^n\Gamma\( \frac{d-1}2 +n\)}{n!\,\Gamma\(\frac{d-1}2\)}\,_3F_2\[\begin{array}{c}
-n, \,2\Delta-d+\frac12, \,2\Delta-d+1 \\
 1-2d +4\Delta,\,\frac32-\frac{d}2-n
 \end{array}
 ; 1  \]\,.
\eea
We can alternatively, re-express those coefficients in terms of Gamma functions as 
\begin{eqnarray}
&&K_\Delta(d,n)=\sum_{k=0}^n\frac{4^{n-k}}{2\,n!} {n \choose k} \frac{\Gamma(4\Delta-2d+2k)}{\Gamma(4\Delta-2 d+k+1)}\nonumber\\
&& \qquad \qquad \qquad \qquad \times\left[\frac{(4\Delta- 2d) \Gamma(\frac{d}{2}+n-k)}{\Gamma(\frac{d}{2}) }+ \frac{(4\Delta-2d+2k)\Gamma(\frac{d-1}{2}+n-k)}{\Gamma(\frac{d-1}{2})}\right]\,,
\end{eqnarray}
which we find more convenient for further manipulations. 
The first three terms in the series expansion (\ref{I-small-x-series}) are given by \footnote{The explicit $K_\Delta(d,n)$ functions for the first five values of $n$ are given by:
\bea
&&K_\Delta(d,0)=1\,,\qquad \qquad \qquad  \qquad \qquad \quad K_\Delta(d,1)=4\,\Delta\,, \nonumber \\
&&K_\Delta(d,2)=d+6\Delta+8\Delta^2\,, \qquad \qquad \qquad K_\Delta(d,3)=\frac{4}{3}(1+\Delta)\(3d+2\Delta(5+4\Delta)\))\,, \nonumber\\
&&K_\Delta(d,4)=\frac{d^2}{2}+\frac{d}{2}\(29+4\Delta\(11+4\Delta\)\)+\frac{\Delta}{3}\(3+2\Delta\)\(5+4\Delta\)\(7+4\Delta\)\,. 
%&&K_\Delta(d,5)=\frac{2}{15}\(2+\Delta\)\[15 d^2+5d\(39+4\Delta\(13+4\Delta\)\)+2\Delta\(3+2\Delta\)\(7+4\Delta\)\(9+4\Delta\)\]\,,
%&&K_\Delta(d,6)=\frac{1}{90}[15 d^3+45d^2\(5+2\Delta\)\(9+4\Delta\)+60d\(281+\Delta\(575+\Delta\(425+ 8\Delta\(17+2\Delta\)\)\)\)   \nonumber\\
%&&\qquad \qquad\qquad \qquad \qquad +4\Delta\(2+\Delta\)\(5+2\Delta\)\(7+4\Delta\)\(9+4\Delta\)\(11+4\Delta\)]\,.
\eea}
\bea\label{I-series}
I^{n\ell}_\Delta(A:B)=\(\frac{z}{4}\)^{2\Delta}\frac{\sqrt{\pi}\Gamma\[2\Delta+1\]}{4\Gamma\[2\Delta+\frac32\]}&+&4\Delta\(\frac{z}{4}\)^{2\Delta+1}\frac{\sqrt{\pi}\,\Gamma\[2\Delta+2\]}{4\Gamma\[2\Delta+\frac52\]}\nonumber\\
&&+(d+6\Delta+8\Delta^2)\(\frac{z}{4}\)^{2\Delta+2}\frac{\sqrt{\pi}\Gamma\[2\Delta+3\]}{4\,\Gamma\[2\Delta+\frac72\]}+\cdots 
\eea
%In appendix {\ref{App-A}} we reproduced the above result following the methods of .... by explicitly computing the contribution to the mutual information from the various replica operators with support in two copies constructed from a single ${\cal O}$ in each copy and their descendants. This represents a non-trivial check of the validity of our formula. 
We will reproduce this formula from (\ref{MI-CB}) by directly computing the first few terms in the long-distance expansion \footnote{Recently, a computation of the mutual information of disjoint spheres for the free massless scalar appears in \cite{Buchanan:2024tlm}, where the first few terms in the long distance expansion were presented. These terms agree exactly with our formula (\ref{I-series}) up to the order at which our formula is exact for the free scalar, $\Delta=(d-2)/2$. This changes with the dimensionality as ${\cal O}((R/r)^{3(d-2)})$, which is determined by the first four copy operator contributions that appears in the twist operator expansion. The methods used in that work were based on \cite{Bramante:2023trx}, and depends strongly on the free character of the theory.}. The operators that contribute to (\ref{int-rep-2}) are simply all the local primary replica operators that can be obtained from ${\cal O}^l(x_1){\cal O}^k(x_2)$ by acting with arbitrary differential operators on $x_1$ and $x_2$ and their associated descendants. In few words, it is the two-copy sector generated by ${\cal O}$. We will evaluate the contributions to $I^{n \ell}_\Delta(A:B)$ from these operators following the methods of \cite{Agon:2015ftl,Chen:2017hbk,Casini:2021raa} and paying attention to the required order in the long-distance expansion.

The leading term in such expansion comes from a replica primary operator of the form $\mathcal{O}_i\mathcal{O}_j$. The contribution of that operator to the mutual information has the well-known form \cite{Agon:2015ftl}
\bea\label{scalar-primary}
\Delta I_{\mathcal{O}^2}= \frac{\sqrt{\pi} \Gamma[2\Delta+1]}{4\Gamma[2\Delta+\frac{3}{2}]}\(\frac{R_A R_B}{L^2}\)^{2\Delta} \,.
\eea
One can obtain the resumed contribution from this replica primary and its descendants by considering the conformal block associated to a scalar operator with scaling dimension $2\Delta$ and normalized according to (\ref{scalar-primary}). This leads to
 \bea
\Delta I^{CB}_{\mathcal{O}^2}=\frac{1}{2^{4\Delta}}\frac{\sqrt{\pi}\Gamma\[2\Delta+1\]}{4\Gamma\[2\Delta+\frac32\]}\(\frac{z^2}{1-z}\)^{\Delta}\,_3F_2\[\begin{array}{c}
 \Delta -\frac{d-2}2,\, \Delta, \, \Delta \\
  2\Delta-\frac{d-2}2, \,\Delta+\frac{1}2
 \end{array}
 ; -\frac{z^2}{4(1-z)}\]\,.
  \eea
 where $z=4R_A R_B/L^2$. To match with (\ref{I-series}) we need the first two subleading terms in the small $z$ expansion of $\Delta I^{CB}_{\mathcal{O}^2}$, these are:
\bea
\Delta I_{\partial_\alpha \mathcal{O}^2}&+&\Delta I_{\partial_\alpha \partial_\beta \mathcal{O}^2}= 4\Delta \frac{\sqrt{\pi} \Gamma[2\Delta+1]}{4\Gamma[2\Delta+\frac{3}{2}]}\(\frac{R_A R_B}{L^2}\)^{2\Delta+1} \nonumber \\
&+&8\Delta\,\frac{(2-d+(8-2d)\Delta+(14-2d)\Delta^2+8\Delta^3)}{(1+2\Delta)(2-d+4\Delta)} \frac{\sqrt{\pi} \Gamma[2\Delta+1]}{4\Gamma[2\Delta+\frac{3}{2}]}\(\frac{R_A R_B}{L^2}\)^{2\Delta+2}\,.
\eea 
The first primary in the replicated theory that we cannot make from descendants of the primary $\mathcal{O}_i \mathcal{O}_j$  is given by $V^{ij}_ \alpha=\mathcal{O}_i \partial_ \alpha \mathcal{O}_j - \partial_ \alpha \mathcal{O}_i\mathcal{O}_j$. Normalizing $\mathcal{O}$ such that 
\begin{equation}
\expval{\mathcal{O}(0) \mathcal{O}(x)} =\frac{1}{\left|x\right|^{2 \Delta}}\,,
\end{equation}
we can easily find
\begin{equation}
\expval{\mathcal{O}(0) \partial_ \alpha \mathcal{O}(x)} =-2 \Delta\frac{ \hat{x}_ \alpha}{\left|x\right|^{2 \Delta +1}}\,, \qquad {\rm and}\qquad \expval{V_ \alpha ( 0) V_ \beta( x)}=  4 \Delta \frac{I_{\alpha \beta}(\hat x)}{\left|x\right|^{4 \Delta +2}}\,,
\end{equation}
with $I_{\alpha \beta}(\hat x)= g_{\alpha \beta} - 2 \hat x_ \alpha \hat x_ \beta$. The contribution to the mutual information is given by
\begin{equation}
\Delta I_{V_ \alpha} =  D n_A^\alpha n_B^\beta \expval{V_ \alpha( 0) V_ \beta (L)} = - \frac{4 \Delta D}{L^{4 \Delta+2}}\,,
\end{equation}
where $n_A$ and $n_B$ are the normal directions to each sphere, and $L$ is the separation between their midpoints. The coefficient $D$ is determined by requiring that the twist operator reduces to modular evolution when inserted into correlation functions, which translates into
\begin{equation}
D= \!\!\! \lim_{|r_{1,2}|\to \infty}\!\!\! \frac{|r_1|^{4 \Delta +2}|r_2|^{4 \Delta +2}}{16 \Delta^2}\!\!\int_{-\infty}^{\infty} \!d s \frac{\pi}{4 \cosh^2 \pi s} \expval{\tilde{\mathcal{O}}_A \partial_ \alpha \mathcal{O}- \partial_ \alpha \tilde{\mathcal{O}}_A \mathcal{O}}(r_1) \expval{\tilde{\mathcal{O}}_B \partial_ \beta \mathcal{O}- \partial_\beta \tilde{\mathcal{O}}_B \mathcal{O}}(r_2) n_A^\alpha n_B^\beta
\end{equation}
where $\tilde{\mathcal{O}}_A$ means the operator $\mathcal{O}$ modularly evolved with parameter $\frac{i}{2}+s$ with the modular hamiltonian of $A$. This is straight-forward to compute and gives
\begin{equation}
D= \frac{\sqrt{\pi}\Delta \Gamma(2 \Delta)}{4 \Gamma(2\Delta +\tfrac{5}{2})} R_A^{2 \Delta +1} R_B^{2 \Delta +1}\,. 
\end{equation}
Then, its contribution to the mutual information becomes
\begin{equation}
\Delta I_{V_ \alpha} = - \frac{2 \Delta}{2 \Delta +1} \frac{\sqrt{\pi} \Gamma( 2 \Delta+ 2)}{4 \Gamma(2 \Delta+ \tfrac{5}{2})} \left(\frac{R_A R_B}{L^2}\right)^{2 \Delta+1}\,.
\end{equation}
To account for the descendants of $V_ \alpha$ we can use the fact that the conformal block contribution for $V_ \alpha$ is given by
\begin{equation}
 \Delta I^{CB}_{V_ \alpha} = \frac{z(z-2) \Delta \sqrt{\pi} \Gamma(1+2 \Delta)}{2^{4 \Delta+4}(z-1) \Gamma(2 \Delta + \tfrac{5}{2})} \left( \frac{z^2}{1-z} \right)^\Delta \,_3F_2 \left(\begin{matrix}
 \Delta +1- \varepsilon, \Delta+1, \, \Delta+1 \\
  \Delta +\tfrac{3}{2}, \,2 \Delta+1- \varepsilon
 \end{matrix}
 ; -\frac{z^2}{4(1-z)}\right)
\end{equation}
where $\varepsilon=\frac{d-2}{2}$. When expanded this gives a contribution for the first descendant
\begin{equation}
\Delta I_{\partial_ \beta V_ \alpha} = - \frac{\sqrt{\pi} \Delta \Gamma(2 \Delta +2)}{\Gamma(2 \Delta+ \tfrac{5}{2})}\(\frac{R_A R_B}{L^2}\)^{2\Delta+2}\,,
\end{equation}
which is independent of $\epsilon$.

For the next contribution, we first have to identify the primary operators. Focusing first on the scalar sector it is possible to show that only
\begin{equation}
V^0=\mathcal{O}_1 \partial^2 \mathcal{O}_ 2 + \mathcal{O}_2 \partial^2 \mathcal{O}_1 - \frac{2 \Delta-d+2}{\Delta} \partial \mathcal{O}_1 \cdot \partial \mathcal{O}_2
\end{equation}
is a primary operator, by checking that its variation under infinitesimal special conformal transformation at $x=0$ vanishes. Its two-point function is given by
\begin{equation}
\expval{V^0 V^0(x)} =\frac{4 d(d-4 \Delta-2)(d-2 \Delta- 2)}{\left|x\right|^{4 \Delta+4}}\,.
\end{equation}
Then, the contribution of this operator can be computed to be
\begin{align}
\Delta I_{V^0}&= \frac{(d-2)^2+(d-2)(3d-4)\Delta+2(3+d(d-2))\Delta^2}{2d (d-2-4 \Delta)}\frac{\sqrt{\pi} (d-2-2 \Delta)\Gamma(2 \Delta+1)}{\Gamma(2 \Delta + \tfrac{7}{2})}\left(\frac{R_A R_B}{L^2}\right)^{2 \Delta+2}\,.
\end{align}
If we consider the operators with two tensor indices, the only combination which transforms as a primary is given by
\begin{equation}
V_{\alpha \beta}= \partial_ \alpha\partial_ \beta \mathcal{O}_1 \mathcal{O}_2 + \mathcal{O}_1 \partial_ \alpha \partial_ \beta \mathcal{O}_2  - \frac{2 \Delta +2}{\Delta} \partial_{(\alpha} \mathcal{O}_1 \partial_{\beta)} \mathcal{O}_2 - \frac{g_{\alpha \beta}}{2} \left( \mathcal{O}_1 \partial^2 \mathcal{O}_2 +\partial^2 \mathcal{O}_1 \mathcal{O}_2 - \frac{2 \Delta+2}{\Delta} \partial \mathcal{O}_ 1 \cdot \partial \mathcal{O}_2 \right) 
\end{equation}
Repeating the same calculation we obtain for this primary
 \begin{align}
 \Delta I _{V_{\alpha \beta}}& =\frac{d-1}{d} \frac{\sqrt{\pi } (\Delta  (3 \Delta +4)+2) \Gamma (2 \Delta +3)}{2 (2 \Delta +1)^2 \Gamma \left(2 \Delta +\frac{7}{2}\right)}\left( \frac{R_A R_B}{L^2} \right)^{2 \Delta+2}\,.
 \end{align}
Summing all the contributions for each order in the long distance expansion we get
\bea
&&\Delta I_{{\cal O}^2}=\frac{\sqrt{\pi} \Gamma[2\Delta+1]}{4\Gamma[2\Delta+\frac{3}{2}]}\(\frac{R_A R_B}{L^2}\)^{2\Delta}\,, \\
&&\Delta I_{\partial_\alpha {\cal O}^2}+\Delta I_{V_\alpha}= 4\Delta\frac{\sqrt{\pi}\,\Gamma\[2\Delta+2\]}{4\Gamma\[2\Delta+\frac52\]}\(\frac{R_A R_B}{L^2}\)^{2\Delta+1}\,, \\
&&\Delta I_{\partial_\alpha \partial_\beta {\cal O}^2}+\Delta I_{V^0} +\Delta I_{V_{\alpha \beta}} +\Delta I_{\partial_ \beta V_ \alpha}=(d+6\Delta+8\Delta^2)\frac{\sqrt{\pi}\Gamma\[2\Delta+3\]}{4\,\Gamma\[2\Delta+\frac72\]}\left( \frac{R_A R_B}{L^2} \right)^{2\Delta+ 2}\!\!.
\eea
In perfect agreement with (\ref{I-series}). This represents a non-trivial check of the validity of (\ref{int-rep-2}). 

\subsection{Universal coefficients and generalized free field \label{UC-GFF}}
The ground state mutual information contains fundamental data about the underlying QFT that can be extracted from the short-distance mutual information of concentric spheres. In general, the mutual information in this regime has the following form
\bea\label{short-d-I}
I=k^{(d)}\frac{{\cal A}(\mathbb{S}^{d-2})}{\delta^{d-2}}+\cdots+2\left\{ \begin{array}{cc} (-1)^{\frac d2-1}4A^{(d)}\log\(R/\delta\) &({\rm even}\,\, d) \\(-1)^{\frac {d-1}2}F^{(d)} & ({\rm odd}\,\, d) 
\end{array}\right.
\eea
where ${\cal A}(\mathbb{S}^{d-2})$ is the area of a sphere of radius $R$, $\delta$ is the radial separation between the concentric spheres, and $k^{(d)}$ is a universal coefficient that characterizes the CFT. In the above expansion, $A^{(d)}$ coincides with the trace-anomaly coefficient corresponding to the Euler density for even-dimensional theories \cite{Duff:1993wm}, while $(-1)^{\frac {d-1}2}F^{(d)}$ equals the free energy of the corresponding CFT evaluated on a round sphere $\mathbb{S}^d$ \cite{Casini:2011kv}. 

\begin{figure}
\begin{center}
\includegraphics[scale=0.45]{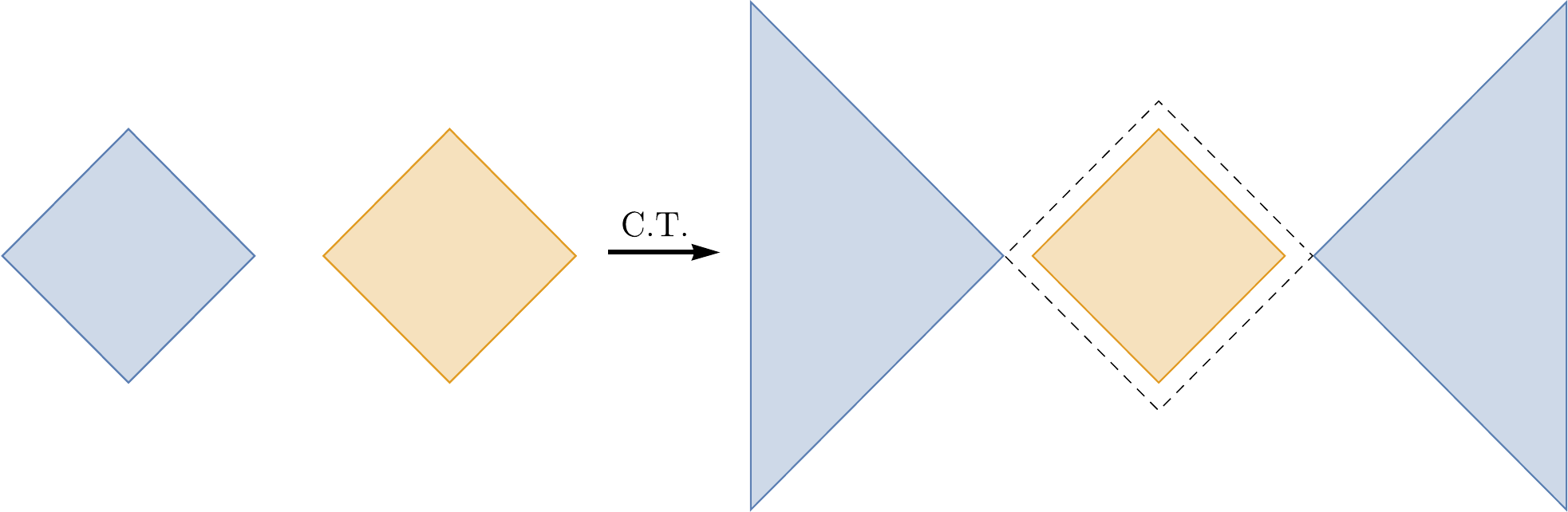}
\begin{picture}(0,0)
\put(-428.5,69.5){\small \circle*{1.5}}
\put(-360.3,69.5){\circle*{1.5}}
\put(-341.2,69.5){\circle*{1.5}}
\put(-273,69.5){\circle*{1.5}}
\put(-156.6,69.5){\circle*{1.5}}
\put(-148.1,69.5){\circle*{1.5}}
\put(-80.4,69.5){\circle*{1.5}}
\put(-72.2,69.5){\circle*{1.5}}
\put(-142.1,68.5){\small $\tilde{x}_B^l$}
\put(-95.4,68.5){\small $\tilde{x}_B^r$}
\put(-335.2,68.5){\small $x_B^l$}
\put(-288,68.5){\small $x_B^r$}
\put(-422.5,68.5){\small $x_A^l$}
\put(-375.3,68.5){\small $x_A^r$}
\put(-66.2,68.5){\small $\tilde{x}_A^l$}
\put(-171.6,68.5){\small $\tilde{x}_A^r$}
\end{picture}
\end{center}
\caption{Graphic illustration of the special conformal transformation that takes the two disjoint spheres $D_A$, and $D_B$ and map them into concentric spheres $\tilde{D}_{A}$, and $\tilde{D}_{B}$. Notice that $\tilde{D}_{A}$ is the space-like complement of an ordinary sphere.}
\label{fig:CT}
\end{figure}
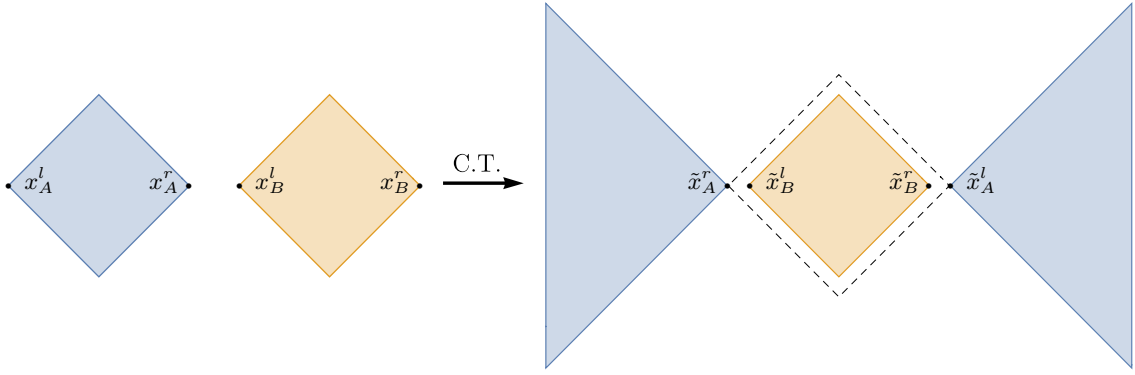

Given the above general structure, we would like to understand the short-distance behavior of $I^{n \ell}_\Delta(A:B)$ for concentric spheres. For disjoint spheres of radius $R_A$ and $R_B$ separated from center to center distance $L$, the conformal ratio $z$ obeys
\bea
z\equiv \frac{|x_1-x_2||x_3-x_4|}{|x_1-x_3||x_2-x_4|}=\frac{4R_AR_B}{L^2-(R_A-R_B)^2}\quad{\rm with }\quad z\in [0,1]\,.
\eea
We take the four points from the intersection between the boundaries of $D_A$ and $D_B$ and the line that joints their centers. This is $x_1=x_A^l$, $x_2=x_A^r$  and $x_3=x_B^l$, $ x_4=x_B^r$. There is a conformal transformation that maps this configuration into one of concentric spheres\footnote{See \cite{Nakaguchi:2014pha} for a detailed discussion of the associated conformal transformation.}. After such transformation the new points become $\tilde{x}_1=\tilde{x}_A^r$, $\tilde{x}_2=\tilde{x}_A^l$  and $\tilde{x}_3=\tilde{x}_B^r$, $ \tilde{x}_4=\tilde{x}_B^l$. 
We illustrate this in figure \ref{fig:CT}. Choosing $\tilde{R}_A=R+\delta$ and $\tilde{R}_B=R$, we get 
\bea
z=\frac{4R(R+\delta)}{(2R+\delta)^2}\underset{\delta \to 0}{\approx} 1-\frac{\delta^2}{4R^2}\,.
\eea
Thus, the limit of coincident concentric spheres corresponds to the $z\to 1^-$ limit of our formulas\footnote{
In the $z\to 1^-$ limit the following relation between Gaussian hypergeometric functions are of utility
\bea\label{zto1-zFid}
\,_2F_1\(a,b,c;z\)&=&\frac{\Gamma(c)\Gamma(c-a-b)}{\Gamma(c-a)\Gamma(c-b)}\,_2F_1\(a,b,a+b+1-c;1-z\)\nonumber\\
&&\qquad +\frac{\Gamma(c)\Gamma(a+b-c)}{\Gamma(a)\Gamma(b)}\(1-z\)^{c-a-b}\,_2F_1\(c-a,c-b,1+c-a-b;1-z\)\,.
\eea
In our case of interest $a=(d-k)/2$, $b=2\Delta+1$ and $c=2\Delta+3/2$.} (\ref{int-rep-2}),(\ref{int-rep-2F1}). 
In such a limit one can write down an expansion in powers of $\delta/2R$ using the identity (\ref{zto1-zFid}). The formulas obtained in this way are easy to analyze since the first few terms control the most divergent terms in the small separation expansion. For instance, the first three terms of the resulting series for arbitrary $d$ are given by:
\bea\label{short-d-2-copy-I}
I^{{n\ell}, d>2}_\Delta(A:B)&=& \frac{\sqrt{\pi}\, \Gamma\(\frac{d-1}2\)}{2^{d+3}\Gamma\(\frac{d}2\)}\(\frac{R}{\delta}\)^{d-1}-
\(\Delta-\frac{d}2\)\frac{\sqrt{\pi}\, \Gamma\(\frac{d-2}2\)}{2^{d+2}\Gamma\(\frac{d-1}2\)}\(\frac{R}{\delta}\)^{d-2} \nonumber\\
&&+\frac{1}{2}\[\(\Delta-\frac{d}2\)^2-\frac{(d-\frac32)(d-1)}{8(d-2)}\]\frac{\sqrt{\pi}\, \Gamma\(\frac{d-3}2\)}{2^{d+1}\Gamma\(\frac{d-2}2\)}\(\frac{R}{\delta}\)^{d-3}+\cdots
\eea
while for $d=2$ we have 
\bea
\label{short-d-2-copy-2d}
I^{{n\ell},d=2}_\Delta(A:B)=\frac{\pi}{32}\(\frac{R}{\delta}\)-\frac{\Delta-1}{8}\log\(\frac{2R}{\delta}\)+\cdots
\eea
Interestingly, we find that the mutual information encoded by our formula has a leading volume divergence. At first sight, this is puzzling since the full mutual information should have an area divergence at short distances as reminded in (\ref{short-d-I}). From a CFT perspective $I^{n\ell}_\Delta(A:B)$ contains the contributions to the mutual information from the two-copy sector of the replica operators generated by the single primary ${\cal O}_{\Delta}$. From (\ref{int-rep-2}) it is clear that this contribution is always positive, so no cancellations can occur among different primaries. Then, the reduction from a volume law to an area law must be related to the contributions of a larger number of replicas which must have necessarily different signs. 
 As we go to the high energy, short distance limit, the contributions of large scaling dimensions and multiple replicas have to be re-summed. We explore this resumation for the single two-copy sector contribution in $d=2$ assuming a Cardy density of states in the Appendix \ref{resumation-2-copy}. We found that the volume divergence at short distances is exponentially enhanced. This observation strengthens the importance of the contributions from the multi-copy sectors in reproducing the expansion (\ref{short-d-I}). On the other hand, one can continue the short distance expansion in (\ref{short-d-2-copy-I}) up to the constant and logarithmic terms to obtain the contribution from $I^{n \ell}_\Delta(A:B)$ to the anomaly coefficients $A^{(d)}$ and $F^{(d)}$ from (\ref{short-d-I}). We do that in Appendix \ref{anomaly-coeffs}, equations (\ref{A-2copy-term}) and (\ref{F-2copy-term}). 

Remarkably, the results (\ref{short-d-2-copy-I}) and (\ref{short-d-2-copy-2d}) have a strong similarity with the mutual information for generalized free fields discussed in \cite{Benedetti:2022aiw}. 
A generalized free field theory (GFF) with conformal symmetry is a ``CFT'' that is determined by a single primary operator with scaling dimension $\Delta$ and Gaussian correlators. No other CFT data enters in the description. It is a peculiar CFT since it does not have a stress tensor. In a holographic description, a GFF is dual to a massive scalar free field theory in Anti-de Sitter space-time \cite{Duetsch:2002hc}.
%However, in a holographic realization of our computation, we expect this contribution to encode the information of the free dual field in an asymptotically Anti-de Sitter space-time.
Various results on the behavior of the MI for GFF can be obtained from this identification. The MI of a GFF for disjoint spheres in $d$ dimensional Minkowski space-time is obtained from the computation of the mutual information 
for concentric hemispheres of a massive free scalar theory in AdS$_{d+1}$ with mass $m^2=\Delta(\Delta-d)$ \cite{Benedetti:2022aiw}. This implies that the MI of a GFF has a leading volume term instead of an area term as $I^{n \ell}_\Delta(A:B)$ does. 
The result of the mutual information of a GFF from \cite{Benedetti:2022aiw} at short distances is  
\bea\label{I-GFF}
I_{GFF}(A:B)=k_{d+1}\frac{\pi^{\frac{d}2}}{\Gamma\(\frac d2\)}\left(\frac{R}{\delta}\right)^{d-1}+
\left\{ 
\begin{array}{cc}
-\frac{\Delta-1}{3}\log\(\frac{R}{\delta}\) +\cdots \qquad \qquad \quad & d=2 \\
-\left(\Delta- \frac{d}{2}\right) \frac{k_d\, \pi^{\frac{d-1}{2}}}{\Gamma\(\frac{d-1}{2}\)}\(\frac{R}{\delta}\)^{d-2}+\cdots & d>2 
\end{array}\right.
\eea
Here, $k_d$ is the coefficient in the area term of the mutual information for a free massless scalar between two planar boundaries in $d$ dimensions \cite{Casini:2009sr, Casini:2005zv}. This short distance expansion has indeed the same form as the expansions (\ref{short-d-2-copy-I}), (\ref{short-d-2-copy-2d}). A numerical comparison using the first few values of $k_d$ shows that the coefficients are also similar. For instance, using $k_3=0.0396506498...,\, k_4=0.0055351600..,\, k_5=0.0013139220...$ for $d=2,3,4$ respectively, gives a difference of the $21\%$, $10\%$, and $5.6\%$ with respect to the volume coefficient in (\ref{short-d-2-copy-I}).
 For higher dimensions, a very good approximation is $k_d\approx \frac{\Gamma\left(\frac{d-2}{2}\right)}{2^{d+2}\pi^{\frac{d-2}{2}}}$. Plugging this approximation into (\ref{I-GFF}) leads to an exact matching between the first two terms in the short distance expansions of $I_{GFF}(A:B)$ and $I^{n \ell}_\Delta(A:B)$.

Of course, the first term in the long-distance expansion also matches since it is determined by the single two-copy primary term (\ref{scalar-primary}). Thus, it is really surprising that (\ref{int-rep-2}) captures exactly the mutual information of the GFF in both the short and long-distance regimes when the number of dimensions $d$ is large, and gives an excellent approximation for lower dimensions. We do not expect $I^{n \ell}_\Delta(A:B)$ to be equal to the mutual information of a GFF, since such mutual information should include contributions from an arbitrary number of replica sectors. Therefore, it would be very interesting to understand the reason behind this matching both for finite $d$ (where it is approximate) and large $d$ (where it is exact). We leave a further exploration of this question to future work.

\section{Mutual information: Euclidean signature}
\label{Euclidean}
In this section, we repeat the calculation of section \ref{MI-Local} and \ref{MI-Multi-local} in Euclidean signature. From a conceptual standpoint, the expansion of the Euclidean twist operator for disjoint regions in terms of local fields is not restricted to a finite region. As a consequence, it does not factor into the product of the twist operators of the individual regions. This implies that our developed framework would not be valid in this context. Nevertheless, and somewhat surprisingly, we will see that we can obtain the right answer by a simple projection operation if one assumes factorization of the twist operators.  Namely, our formulas will give rise to the right result plus extra pieces associated with the shadow operators.

\subsection{Mutual information: Local kernel} 
We consider the analog local kernel expansion of (\ref{local-ansatz}) in Euclidean signature. The local ansatz in this case involves an integral over the whole spacetime and the action of the kernel is defined through  
\bea\label{Eucl-kernel-coor-mod}
 \sum_{p<q}\int d^d\xi\,\tilde{\cal G}^{pq}_A(\xi) \langle {\cal O}^p(\xi){\cal O}^q(\xi){\cal O}^l(x){\cal O}^k(x)\rangle &=\langle{ \cal O}(x){\cal O}_{A}\[x, \theta_{lk}\]\rangle\,,
\eea
where $\theta_{lk}\equiv 2\pi(l-k)/n$.
The correlator on the right-hand side of (\ref{Eucl-kernel-coor-mod}) is obtained with respect to the Euclidean modular flow:
\bea\label{Eucl-sphere-modular}
x^{\pm}_A[x,\theta]=R_A\frac{\(R_A+x^{\pm}\)-e^{\mp i \theta} \(R_A-x^{\pm}\) }{\(R_A+x^{\pm}\)+e^{\mp i \theta} \(R_A-x^{\pm}\)}\, ,
\eea
where $x^\pm=r\pm i \tau$ are standard null coordinates in Euclidean signature. This transformation follows from (\ref{sphere-modular}) by $s\to i \theta/2\pi$. The associated correlator transforms as in (\ref{modular-corr}). However, in this case, the Jacobian factor is given by 
 \bea\label{Jacobian-Eucl}
 \left|\frac{\partial x_{2;A}(\theta)}{\partial x_2 }\right|=\frac{2R_A^2}{r^2+\tau^2+R_A^2-(r^2+\tau^2-R_A^2)\cos\theta -2R_A \tau \sin \theta}\,.
 \eea
Using this, the expression (\ref{Eucl-kernel-coor-mod}) simplifies to 
\bea
\label{Eucl-kernel-corr-mod-2}
\int d^d\xi\,\tilde{\cal G}^{kl}_A(\xi) \frac{1}{|x-\xi|^{4\Delta}}= \frac{1}{4^\Delta \sin^{2\Delta}\( \frac{\theta_{lk}}{2}\)}\frac{\(2R_A\)^{2\Delta}}{\[(r-R_A)^2+\tau^2\]^{\Delta}\[ (r+R_A)^2+\tau^2\]^{\Delta}}\,.
\eea
Before proceeding, let us notice two main differences with the Lorentzian formula. First, the modular Euclidean evolution is evaluated on real values of the modular parameter. This means that we can make sense of $\tilde{\cal G}^{kl}_A(\xi)$ before the analytic continuation of the sum over $\{k,l\}$ to an integral over $s$ used in section \ref{MI-Local}. Second, the integral over the kernel is performed on the whole of the space-time as opposed to the causal diamond, as it was done in section \ref{MI-Local}. As in the Lorentzian case, we can identify the above equation with (\ref{Eucl-inverse-CB}), and read off from it the Kernel:
\bea
\tilde{\cal G}^{kl}_A(\xi)=\frac{c_{2\Delta}}{4^\Delta \sin^{2\Delta}\( \frac{\theta_{lk}}{2}\)}\[\frac{\(|\vec{x}|^2+\tau^2-R_A^2\)^2+4\tau^2 R_A^2}{(2R_A)^2}\]^{\frac{d}{2}-\Delta}\,,
\eea
where $c_{2\Delta}$ is given by (\ref{Eucl-CB-coeff}). The contribution to the Euclidean twist operator of this replica primary is thus given by 
\bea
\tilde{\Sigma}_A^{(n)}\approx \sum_{k<l}\frac{1}{4^\Delta \sin^{2\Delta}\( \frac{\theta_{lk}}{2}\)}\left\{c_{2\Delta}\mathlarger{\int} \[\frac{\(|\vec{x}|^2+\tau^2-R_A^2\)^2+4\tau^2 R_A^2}{(2R_A)^2}\]^{\frac{d}{2}-\Delta} {\cal O}^l(\xi){\cal O}^k(\xi)d^d\xi\right\}
\eea
This leads to an expansion of the twist operator in terms of OPE block-like objects (see (\ref{euc-rep}) for a comment on the difference between this and the actual OPE block). Plugging this into the formula for the mutual information results in an expansion of the mutual information in terms of conformal block-like objects 
\bea\label{Eucl-Mutual-CB-scalar}
\tilde{I}^{\ell}_\Delta(A:B)=\lim_{n\to 1}\frac{1}{n-1}\sum_{k<l}\frac{1}{2^{4\Delta} \sin^{4\Delta}\( \tau_{lk}\)}\tilde{G}_{2\Delta}(u,v)=\frac{1}{2^{4\Delta}}\frac{\sqrt{\pi} \Gamma[2\Delta+1]}{4\Gamma[2\Delta+\frac{3}{2}]}\tilde{G}_{2\Delta}(u,v)
\eea
where the basis of functions $\tilde{G}_{2\Delta}(u,v)$ corresponds to a linear combination of the conformal block associated with an operator of dimension $2\Delta$ and its shadow partner of dimension $d-2\Delta$ \cite{Dolan:2000ut,Fitzpatrick:2011hu,SimmonsDuffin:2012uy}. After projecting out the shadow contribution, we arrive at the general expansion of the mutual information 
\bea\label{Eucl-MI-CB}
I(A:B)=\sum_{\Delta,J}b_{\Delta,J} G^d_{\Delta,J}(u,v)\,,
\eea 
in terms of conformal blocks as that of the Lorentzian case. 

\subsection{Mutual information: Multi-local kernel}
For the multi-local case we start from the same ansatz as before, adapted to Euclidean signature. In this case, there is no causal diamond so we expect the integral to be over the full space. Apart from that, the Lorentzian analysis goes through unchanged, and using the inverse correlator we find:
\bea
 {\mathcal G}_W(\xi_1, \xi_2;\theta)&=&\int d^d w_1 G^{-1}_\Delta (\xi_1,w_1)
 \left|\frac{\partial w_1(\theta)}{\partial w_1}\right|^{\Delta} \delta^{d}\(\xi_2-w_1(\theta)\)\,,
\\
 {\mathcal G}_W(\chi_1, \chi_2;\theta)&=&\int d^d \zeta_2 G^{-1}_\Delta (\chi_2,\zeta_2)
 \left|\frac{\partial \zeta_2(-\theta)}{\partial \zeta_2}\right|^{\Delta} \delta^{d}\(\chi_1-\zeta_2(-\theta)\)\nonumber\,.
\eea
Both equations are equivalent, as follows from  the identity $\langle \cO_A[w_1,\theta]  \cO(w_2) \rangle=\langle \cO(w_1) \cO_A[w_2,-\theta]   \rangle$, just as in the Lorentzian case. Note that in this case the inverse propagators are just given by a correlator with the appropiate conformal dimension and we do not need to conjecture their existence as in the Lorentzian calculation. The mutual information is given by 
\bea
\tilde{I}^{n \ell}_\Delta(A:B)= \sum_{\theta}\int d^d \xi_1 \int d^d \xi_2\, \int d^d \chi_1 \int d^d \chi_2\,\frac{{\mathcal G}_W(\xi_1, \xi_2;\theta)\,{\mathcal G}_W(\chi_1, \chi_2;\theta)}{\left| \xi_1-\chi_1-L\right|^{2\Delta}\left| \xi_2-\chi_2-L\right|^{2\Delta}} 
\eea
where $L$ is the vector that goes from the center of $A$ to the center of $B$ and we are using the notation:
\bea
\lim_{n\to 1}\frac{1}{n-1}\sum_{j=1}^{n-1}=\sum_\theta\,.
\eea
The same manipulations as in the Lorentzian case lead us to
 \bea
\tilde{I}^{n \ell}_{\Delta}(A:B)=\sum_{\theta} \sum_{i}
\left|\frac{\partial w_1(\theta)}{\partial w_1}\right|_{w=w_i^*}^{\Delta} 
 \left|\frac{\partial \zeta_2(-\theta)}{\partial \zeta_2}\right|^{\Delta}_{\zeta_2=w_1(\theta)-L} \,\,\Bigg|_{w=w_i^*}
\frac{1}{\left|\det \(\frac{\partial f^\alpha(w)}{\partial w^\beta}\)\right|_{w=w_i^*}} \,,
\eea
where $w_i^*$ are the zeros of the function $f(w)$ defined as 
\bea
\label{eq:zeros_euclidean}
f(w)=w-L-x_B\[w(\theta)-L,-\theta\]\,.
\eea
Note that the theta functions are now absent since the integrals are over the full space-time.
The solutions to the above equation come in pairs and are described by two different functions depending on sign of the combination $\sin^2\(\frac{\theta}{2}\)-z$ where $z=4R^2/L^2$.  
\bea
\tilde{w}^*_{1,\pm}&=\left\{-\dfrac{L\sqrt{z}}{2}\cot\(\frac{\theta}{2}\)\pm \dfrac{L\sqrt{z-\sin^2\(\frac{\theta}{2}\)}}{2 \sin\(\frac{\theta}{2}\)},\dfrac{L}{2}\right\} \qquad \text{if}\quad \sin^2 \left(\frac{\theta}{2}\right)<z\,,\\
\tilde{w}^*_{2,\pm}&=\left\{-\dfrac{L\sqrt{z}}{2}\cot\(\frac{\theta}{2}\),\dfrac{L}{2}\pm \dfrac{L\sqrt{\sin^2\(\frac{\theta}{2}\)-z}}{2 \sin\(\frac{\theta}{2}\)}\right\} \qquad \text{if} \quad\sin^2 \left(\frac{\theta}{2}\right)>z\,.
\eea
These are precisely the Wick rotations of the Lorentzian solutions. In the Lorentzian case, only $w^*_{2,+}$ contributed to the mutual information, however, in the Euclidean case there is no reason to choose any particular solution. In particular, this means that $w^*_{2,-}$ must be included giving a contribution that corresponds to the shadow transform of the operator $\mathcal{O}$. This is a known issue in Euclidean signature and to obtain the right result one must remove the shadow contribution by hand.
%Interestingly this set of points lies on a circle obeying
%\bea
%\(\tilde{w}^*_{r\pm}-\frac{1}{2}\)^2+\tilde{w}^{*2}_t=\(\frac{1-z}{4}\)
%\eea
%Notice that the circle where the solutions lies is center at the middle point between spheres $r=1/2$, in our rescaled coordinates, and it always links the two spheres since
%\bea
%\tilde{w}_{r_{min}}=\frac{1}{2}-\frac{\sqrt{1-z}}{2}, \qquad {\rm and }\qquad  \tilde{w}_{r_{max}}=\frac{1}{2}+\frac{\sqrt{1-z}}{2}
%\eea
%satisfies 
%\bea
%0<\tilde{w}_{r_{min}}<\frac{\sqrt{z}}{2}\qquad {\rm and}\qquad 1-\frac{\sqrt{z}}{2}<\tilde{w}_{r_{max}}<1\,.
%\eea
%\begin{itemize}
%\item For $\sin^2\(\frac{\theta}{2}\)-z<0$ 
%\end{itemize}
%\bea
%\tilde{w}^*_{r\pm}=\frac{1}{2}\,, \qquad \tilde{w}^*_{t\pm}=-\frac{x}{2}\cot\(\frac{\theta}{2}\) \pm \frac{\sqrt{x^2-\sin^2\(\frac{\theta}{2}\)}}{2 \sin\(\frac{\theta}{2}\)}
%\eea
%The geometry of these critical points differ considerably from the previous ones as these lie on the line $\tilde{w}_r=1/2$. However, for fixed $x$ these two curves connect as a function of the angle in a continuos although non smooth way.  
The computation of the Jacobian is detailed in appendix \ref{det-euclidean} giving the result for the mutual information
\bea
\tilde{I}^{n \ell}_{\Delta}(A:B)= \sum_{\theta} &&\Bigg[z^{2\Delta}\frac{\(\sin\(\frac{\theta}{2}\)+ \sqrt{\sin^2\(\frac{\theta}{2}\)-z}\)^{2(d-2\Delta)}}{4^d\sin^d\(\frac{\theta}{2}\)\(\sin^2\(\frac{\theta}{2}\)-z\)^{\frac{d}2}}+\nonumber \\
&&\quad +z^{2(d-\Delta)}\frac{\(\sin\(\frac{\theta}{2}\)+ \sqrt{\sin^2\(\frac{\theta}{2}\)-z}\)^{2(2\Delta-d)}}{4^d\sin^d\(\frac{\theta}{2}\)\(\sin^2\(\frac{\theta}{2}\)-z\)^{\frac{d}2}}\Bigg]\,,
\eea
where the sum is over angles with $\sin^2\left(\frac{\theta}{2}\right)-z>0$. Finally we want to project out the shadow contribution and do the replacement:
\bea
\frac{1}{n-1}\sum_{i<j} \to \frac{\pi}{2}\,\int_0^\infty \frac{ds}{\cosh^2\pi s}\,.
\eea
After the analytic continuation, since $\cosh\pi s>1$, we have a single integral expression for that contribution regardless of the values of the angles:
\bea
\tilde{I}^{n \ell}_\Delta(A:B)&=&\frac{z^{2\Delta}}{4^{d+1}}\,\int_{0}^\infty \frac{2\pi\, ds}{\cosh^{d+2}\pi s} \frac{\left(  \cosh \pi s + \sqrt{\cosh^2\pi s-z}\right)^{2(d-2\Delta)}}
{ \(\cosh^2\pi s-z \)^{\frac{d}2}  } \,.
\eea
In this way, we reproduce the same formula as in the Lorentzian case. 

\subsection{Euclidean vs Lorentzian}
By studying more in depth the structure of solutions to \eqref{zeros-g} and \eqref{eq:zeros_euclidean} we can understand which solutions need to be kept in Euclidean signature and how the shadow contributions disappear in Lorentzian signature. For this reason it is convenient to define the matrix 
 \begin{equation}\label{M}
M(s):=\left.\frac{\partial x^J_B[x_1,-s]}{\partial  x_1 } \cdot \frac{\partial x^J_A[x_2,s]}{\partial x_2 } \right|_{w_i^*} \,,
\end{equation}
where the evaluation stands for $x_2=w_i^*$ and $x_1=x^J_{A}[w_i^*,s]-L$\, and we are working in Lorentzian signature. In terms of $M(s)$, the mutual information takes the simple form:
\begin{equation}\label{mutual-bi-local-6}
I^{n \ell}_\Delta(A:B)=\frac{1}{4}\int_{-\infty}^\infty \frac{\pi \,ds}{\cosh^2\pi s}\sum_ i  \frac{\left|\det M(s)\right|^{\frac{\Delta}{d}}}{\left|\det(1- M(s))\right| }\Theta\(x_{A}^J[w_i^*,s]-L\in D_{\bar{B}} \)\Theta\(w_i^*-L\in D_{B} \)\,.
\end{equation}
From (\ref{M}) and the composition property of the modular flow in the form $x^J_{A/B}[x^J_{A/B}[w,\pm s],\mp s]=w$ and the property $x_{A/B}^J(-w,s)=-x_{A/B}^J(w,-s)$, one can easily find a convenient expression for the inverse of $M$, namely
\bea\label{inv-M}
M^{-1}(s)=\left.\frac{\partial x^J_A[x_1,s]}{\partial  x_1 }\cdot  \frac{\partial x^J_B[x_2,-s]}{\partial x_2 } \right|_{L-w_i^*}\,,
\eea
where in this case the evaluation stands for $x_2=L-w_i^*$, and $x_1=x^J_B[L-w_i^*,-s]-L$. We used the equation for the zeros of (\ref{zeros-g}) to simplify the arguments of the above expression. 
The following relation between $M$ and $M^{-1}$ follows from (\ref{inv-M}),
\bea
M^{-1}(w_i^*(s);A,B,s)=M(L-w_i^*(s);B,A,-s)\,.
\eea
For simplicity, we will restrict ourselves to the case in which the spheres have equal radii $R$, and since we use coordinates adapted to the spherical regions, this implies $A=B$. Note that this situation is always reachable by a conformal transformation. In that case, we obtain
\begin{equation}
\label{eq:relation_shadow}
\frac{\left|\det M(w_i^*(s),s)\right|^{\frac{\Delta}{d}}}{\left|\det(1- M(w_i^*(s),s))\right| } = 
\frac{\left|\det M(L-w_i^*(s),-s)\right|^{\frac{d-\Delta}{d}}}{\left|\det(1- M(L-w_i^*(s),-s))\right| } \,.
\end{equation}
One can further notice that the solutions $w_{i,\pm}$ fulfill the relation
\begin{equation}
 w^*_{i,\pm}(s)= L -w^*_{i,\mp}(-s)\,.
 \end{equation}
When combined with \eqref{eq:relation_shadow} this immediately tells us that
\begin{equation}
\frac{\left|\det M(w_{i,\pm}^*(s),s)\right|^{\frac{\Delta}{d}}}{\left|\det(1- M(w_{i,\pm}^*(s),s))\right| }  = 
\frac{\left|\det M(w_{i,\mp}^*(-s),-s)\right|^{\frac{d-\Delta}{d}}}{\left|\det(1- M(w_{i,\mp}^*(-s),-s))\right| }\,.
\end{equation}
Thus, we find that the solutions $w^*_{2,-}(s)$ are precisely the shadow contributions that are removed in Lorentzian signature by the theta functions and need to be manually projected out in Euclidean signature.

\section{Discussion and future work} 
\label{conclusion}

In this work, we introduced a novel methodology for computing mutual information in CFTs. The key technical advancement was the development of a technique that resums the contributions from all replica operators generated by a single scalar primary supported on two replicas known as the two copy sector, into a single term characterized by that primary. We applied this technique specifically to the case of two disjoint spheres, yielding a straightforward expression for its contribution to the mutual information (\ref{int-rep-2}). This advancement significantly brings us closer to the possibility of computing the full mutual information for arbitrary CFTs in spherical regions.

While this work represents a significant step forward, two important generalizations are still necessary to compute the full mutual information of an arbitrary CFT. These are:

\begin{itemize}
\item \underline{N-copy sector}: Our formula includes only the two-copy sector contribution to the mutual information from a single primary. Therefore, it is essential to extend our framework to accommodate an arbitrary number of copies. This is especially important given the approximate match between (\ref{int-rep-2}) and the mutual information of a generalized free field. These issues will be addressed in forthcoming work by the authors.

\item \underline{Other spin operators}: 
To compute the mutual information for an arbitrary CFT, one needs to derive analogous formulas to (\ref{int-rep-2}) that account for the contributions from primary operators of all possible spins \cite{Casini:2021raa}. Additionally, these formulas must be extended to include contributions from the N-copy sector, as mentioned in the previous item.

\end{itemize}
Once, we have developed the framework to compute the contributions of arbitrary spin primary operators both in the two and N-copy sectors, the full mutual information will be given as an infinite sum over all the primary operators in the theory and over all the possible number of copy sectors.
Besides this goal, other interesting research directions include:
\begin{itemize}
\item \underline{Holography}: It would be highly valuable to identify the holographic duals of the boundary contributions outlined in our framework and develop the appropriate technology to compute them. This endeavor could be viewed as an operator-level refinement of the Faulkner-Lewkowycz-Maldacena proposal  \cite{Faulkner:2013ana} applied to holographic mutual information \cite{Agon:2015ftl}.

\item \underline{Other information theoretic observables}: Many quantum information observables can be calculated within the replica trick framework \cite{Lashkari:2014yva,Calabrese:2012ew,Dutta:2019gen,Yin:2022toc}. Similarly, many of these quantities can be re-expressed in terms of expectation values of twist operators, making the methods developed in this work applicable to a broader range of scenarios \cite{Sarosi:2016atx,Agon:2020fqs,Ugajin:2016opf}. 

\end{itemize}
At a more refined conceptual level, certain issues encountered in this work warrant special attention and may require further investigation. One of the most pressing concerns is the understanding of our framework in the Euclidean signature. Notably, most advancements in computing entanglement measures in QFTs have been achieved within the context of Euclidean QFTs. However, the operator-based approach presented in this work relies heavily on the Minkowski signature and the space-time localization of replica twist operators. Specifically, in Minkowski space, the twist operator for disjoint regions is the direct product of the twist operators for the individual regions a property that does not hold in Euclidean space. This discrepancy calls for a deeper exploration of twist operators in Euclidean signature and the corresponding ``analytic continuation'' needed to translate Lorentzian objects to Euclidean ones and vice versa within this framework.

The localization aspects mentioned above play a crucial role in the evaluations presented in Section \ref{Euclidean}. Although we concluded that simply removing the shadow part from the Euclidean two-copy sector contribution to the mutual information gives the right result, the rationale behind the validity of this method requires further understanding. Another theme, which has implications beyond the scope of this paper, concerns the nature of both OPE blocks and conformal blocks in Euclidean signature. Specifically, are Euclidean conformal blocks equivalent to their Lorentzian counterparts? Does the relationship
\bea
\tilde{G}_\Delta(u,v)=\langle \tilde{\cal B}_\Delta(x_1,x_2)\tilde{\cal B}_\Delta(x_3,x_4)\rangle
\eea
holds for the Euclidean OPE blocks that appear in (\ref{OPE-Eucl})? and finally, Is our Euclidean OPE block representation (\ref{OPE-Eucl}) equivalent to the more familiar ones \cite{SimmonsDuffin:2012uy}?

Finally, it is worth noting that the assumption of analyticity in the integer indices of the multiple integral expression (\ref{mutual-lorentzian-1}) could fail under certain circumstances. However, in the present context, this assumption is justified by the finiteness of the results obtained.

\section*{Acknowledgements}
H.C. and C.A. thank useful discussions with Gonzalo Torroba in the earlier stages of the project. C.A. also acknowledges useful discussions with Murat Kologlu and Kamran Salehi. 
The work of H.C is partially supported by CONICET, CNEA and Universidad Nacional de Cuyo, Argentina. The work of U.G., C.A., and G.P. is supported by the Netherlands Organisation for Scientific Research (NWO) under the VICI grant VI.C.202.104.
 
\appendix 
\section{The Modular Conjugation Operator \label{Tomita-J}}
In this section we analyze the action of the Tomita operator $J$. Following the expression \eqref{sphere-modular} with $s=\frac{i}{2}$, one obtains that the action of $J$ for a sphere is given by
\begin{equation}\label{J-action-1}
  x^0\to  -\frac{R^2 x^0}{|x|^2}\,, \quad {\rm and }\quad  x^i\to  \frac{R^2 x^i}{|x|^2}\,,
\end{equation}
this is the product of an inversion and a time reversal transformation. Under this transformation, the causal diamond of the sphere is mapped not only into the causal diamond of the complement as it should, but also onto the future and past light-cones of the tips of the causal diamond\footnote{ In some circumstances, one can argue that the local map \eqref{J-action-1} represents the action of the Tomita operator in all of Minkowski space. This happens in the free massless field case considered, for instance, in section V of \cite{Haag:1992hx}. For free massless fields, operators commute both at space-like and time-like separations, with non-trivial commutation relations only between light-like separated operators. Due to this peculiar feature, one can include in the commutant of the algebra associated to the causal diamond not only the region space-like separated from it but also the time-like separated regions that lie at the future and past of the tips of the causal diamond. For interacting theories, this is no-longer the case, and the above assignment is not possible.}.

 This inconsistency is related to the fact that Minkowski space does not fully cover the conformal cylinder where the CFT  actually lives  \cite{Segal1971, Segal1976, Luscher:1974ez}. As a result, we can only trust the above local map on points that are mapped smoothly and include the $t=0$ hyper-surface (where the transformation is a simple spatial inversion). This corresponds to the set of points that are space-like separated from the center of the sphere. For points that are time-like separated, the above local map is not valid anymore as they are separated by a singular surface under the map. A recent discussion on the limitations of the sphere modular flow described in \eqref{sphere-modular} can be found in section two of \cite{Casini:2017roe}. As pointed out there, a point in the spatial complement of the sphere under the modular flow \eqref{sphere-modular} goes to infinity at a finite value of the modular parameter, making the mapping for larges values of the modular parameter non-local and no-longer described by \eqref{sphere-modular}. On the other hand, the same map is local for all values of the modular parameter in the cylinder. As the authors concluded, one can find the non-local map in Minkowski for larger values of the modular parameter by propagating the field data in the cylinder back to the $t=0$ slice and mapping that information back to Minkowski. The same strategy will apply to the Tomita map. 

To correctly account for conformal transformations one must compactify Minkowski space, which is naturally done through an embedding into $\mathbb{R}^{2,d}$ \cite{Weinberg:2012mz}. In particular, a point in compactified Minkowski space must be identified with a ray of points $P\in \mathbb{R}^{2,d}$ with $|P|^2=0$ quotiened by the relation $P\sim \lambda P$ for $\lambda\in \mathbb{R}$. Under this description, the Minkowski patch is covered by the slice of the light-cone in $\mathbb{R}^{2,d}$ parametrized as
\begin{equation}
(P^+,P^- ,P^\mu) = (1, |x|^2 , x^\mu)
\end{equation}
where the first two components describe light-cone coordinates along the extra directions that we have added to construct $\mathbb{R}^{2,d}$. This indeed covers all of Minkowski space, only contains a single representative of the quotient characterized by the condition $P^+=1$, and has $|P|^2=0$. However, there are some points in the conformal cylinder which are not included in this parametrization, namely, those with $P^+=0$\footnote{See \cite{Simons-LCFT}, for a recent review of the embedding space formalism in Lorentzian signature}. The fact that we are missing these points is what makes the naive $J$ operator \eqref{J-action-1} to act incorrectly in some regions of the Minkowski patch.

\begin{figure}
\begin{center}
\includegraphics[scale=0.55]{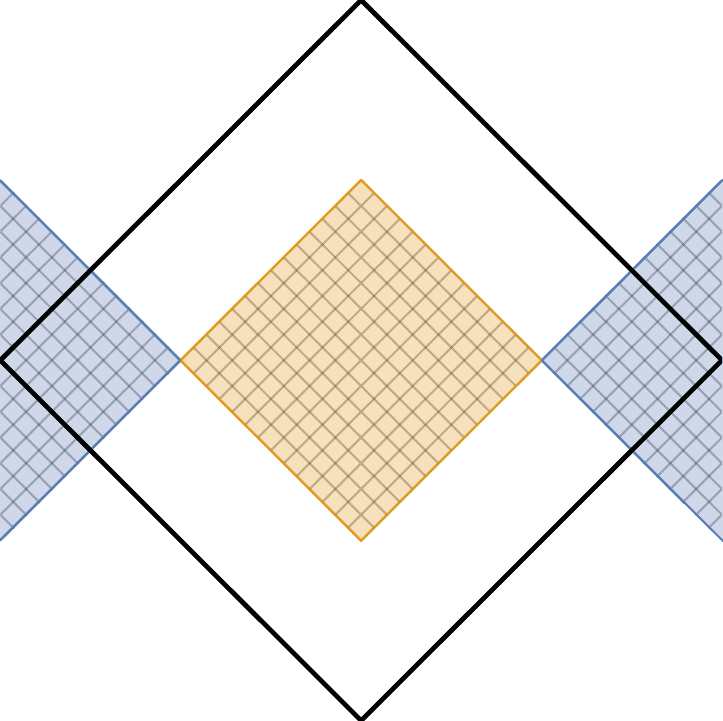}
\end{center}
\caption{Action of the $J$ operator on the conformal cylinder. In orange the original sphere, in blue the image of the sphere under the $J$ operator and in black the boundary of the Minkowski patch.}
\label{fig:J}
\end{figure}

If we want to describe the action of $J$ we need to use a global parametrization of the conformal cylinder. A convenient choice is
\begin{equation}\label{global-cylinder}
(P^+,P^- ,P^0,P^i) = (\cos \sigma+  n^d, \cos \sigma - n^d,\sin \sigma, n^i)\,,
\end{equation}
with $(n^i,n^d)\in \mathbb{R}^d$ with $|n|^2=1$ and $\sigma\in S^1$. This parametrizes $S^1\times S^{d-1}$. Thus, to avoid issues with causality we need to decompactify the time-like $S^1$ by taking its universal cover. Notice that the cylinder section in the embedding space is characterized by the condition 
\bea\label{cylinder-section}
(P^{-1})^2+(P^d
)^2+P^\mu P^\mu=2\,,
\eea
where $P^{-1}$ and $P^d$ are the additional time-like and space-like coordinates respectively. These are related to the null coordinates through $P^{\pm}\equiv P^{-1}\pm P^{d}$.
One can easily find the embedding space matrix that implements the action of $J$ as described by \eqref{J-action-1}, this is
 \begin{equation}
  M_J= \begin{pmatrix}
    0 & \frac{1}{R^2} &0\\
    R^2 &0&0\\
    0&0 &\mathbb{I} 
  \end{pmatrix}\,.
\end{equation}
Remembering the identification $P\sim \lambda P$ this does indeed reproduces \eqref{J-action-1}. However, using the global patch \eqref{global-cylinder} one can find the associated map $J$ on the cylinder by applying $M_J$ to \eqref{global-cylinder} and re-scaling the result to stay within the gauge \eqref{cylinder-section}. The result is 
\begin{equation}
(\cos \sigma+  n^d, \cos \sigma - n^d,\sin \sigma, n^i)\to \frac{2\left(  \frac{\cos \sigma-n^d}{R^2},R^2(n^d+\cos \sigma),-\sin \sigma, n^i \right) }{\sqrt{4\sin^2 \sigma+\frac{(n^d(R^4-1)+(R^4+ 1)\cos \sigma)^2}{R^4}}} \,,
\end{equation}
which represents also a special conformal transformation times a time reflection. The above map is well defined everywhere in the cylinder, as it can be seen from the fact that the denominator is positive definite. We plot this map in Figure \ref{fig:J}, where we can see explicitly that the map exchange the causal diamond of the sphere with the causal diamond of its spatial complement. This also shows that part of the causal diamond of the complement lies outside of the Minkowski patch, and represents precisely the part of the image of the map that is represented incorrectly in the Minkowski section. 

\section{Conformal blocks and OPE blocks \label{OPE-blocks}}

\subsection{Lorentzian representation \label{mink-rep}}
Consider the following integral representation of the OPE block associated to the region $A$ for a scalar operator $\mathcal{O}_\Delta$, \cite{Czech:2016xec, SimmonsDuffin:2012uy} 
\bea\label{OPE-int-rep-6}
{\mathcal{B}}_\Delta (x_A^+,x_A^-)=\tilde{c}_\Delta \int_{D_A}\diff^d \xi \(\frac{|x_A^+-\xi ||x_A^--\xi |}{|x_A^+-x_A^-|}\)^{\Delta-d}\mathcal{O}_\Delta(\xi )\,.
\eea 
For simplicity let us take the points $x_A^+=(R_A,\vec{0})$, and $x_A^-=(-R_A,\vec{0})$.
The Kernel in the above integral has an interesting geometric interpretation, it is given by $|K|^{\Delta-d}$ where $|K|$ is the norm of the conformal killing vector that preserves the causal diamond $D_A$. This conformal killing vector generates the modular flow associated to $D_A$. 

We normalize the OPE block such that in the small $|x_A^+-x_A^-|$ limit, 
\bea \label{normalizationOPE}
{\mathcal{B}}_\Delta(x_A^+,x_A^-)\approx |x_A^+-x_A^-|^{\Delta} \mathcal{O}_\Delta(\bar{\xi} )= (2R_A)^{\Delta} \mathcal{O}_\Delta(\bar{\xi} )\,,
\eea
where $\bar{\xi}$ is the center of the sphere $A$. The normalization coefficient $\tilde{c}_{\Delta}$ can be easily computed. In the $R_A\to 0$ limit, we can pull the operator $\mathcal{O}_\Delta({\xi} )$ out of the integrand (\ref{OPE-int-rep-6}), as 
\bea
{\mathcal{B}}_\Delta(x_A^+,x_A^-)\approx \tilde{c}_\Delta\, \mathcal{O}_\Delta(\bar{\xi} ) \int_{D_A}\diff^d \xi \(\frac{|x_A^+-\xi ||x_A^--\xi |}{|x_A^+-x_A^-|}\)^{\Delta-d}\,,
\eea 
and evaluate the remaining integral\footnote{
To evaluate the integral we use null-like coordinates and carry out first the angular integrals which results in 
\bea
 &&\int_{D_A}\diff^d \xi \(\frac{|x_A^+-\xi ||x_A^--\xi |}{|x_A^+-x_A^-|}\)^{\Delta-d}
 %\nonumber \\
 %&&\qquad \qquad 
 =\frac{|x_A^+-x_A^-|^{d-\Delta}}{\pi^{\frac{1-d}2}\Gamma\(\frac{d-1}2\)}\int d\xi_+ \int d\xi_- \(\frac{\xi_++\xi_-}{2}\)^{d-2} \(R^2_A-\xi_+^2\)^{\frac{\Delta-d}2}\(R^2_A-\xi_-^2\)^{\frac{\Delta-d}2}\,,\nonumber
 \eea
where we used $S_{d-2}=2\pi^{\frac{d-1}2}/\Gamma\left(\frac{d-1}2\right)$. The remaining integral can also be carried out explicitly. By carefully considering the constraint $\xi_++\xi_-\geq 0$, we find 
 \bea
 &&\int d\xi_+ \int d\xi_- \(\frac{\xi_++\xi_-}{2}\)^{d-2} \(R^2_A-\xi_+^2\)^{\frac{\Delta-d}2}\!\!\!\! \(R^2_A-\xi_-^2\)^{\frac{\Delta-d}2}= \frac{R_A^{2\Delta-d} \Gamma\(\frac{d-1}{2}\) \Gamma\(\frac\Delta2\) \[\Gamma\(\frac{\Delta+2-d}{2}\)\]^2}{2^{d-\Delta}\Gamma\(\frac{\Delta+1}2\)\Gamma\(\Delta+1-\frac{d}2\)}\,,
 \eea
 where $R_A=|x_A^+-x_A^-|/2$.} 
 \bea
 \int_{D_A}\diff^d \xi \(\frac{|x_A^+-\xi ||x_A^--\xi |}{|x_A^+-x_A^-|}\)^{\Delta-d}=\frac{\pi^{\frac{d-1}2}\Gamma\(\frac\Delta2\) \[\Gamma\(\frac{\Delta+2-d}{2}\)\]^2}{\Gamma\(\frac{\Delta+1}2\)\Gamma\(\Delta+1-\frac{d}2\)}R_A^{\Delta}\,
 \eea
which implies 
\bea\label{coeff-calpha}
\tilde{c}_\Delta=\frac{2^{\Delta}\Gamma\(\frac{\Delta+1}2\)\Gamma\(\Delta+1-\frac{d}2\)}{\pi^{\frac{d-1}2}\Gamma\(\frac\Delta2\) \[\Gamma\(\frac{\Delta+2-d}{2}\)\]^2}\,.
\eea
The conformal block associated to the intermediate operator ${\cal O}_\Delta$ is given by 
\bea
G_{\Delta}(z)=\langle {\mathcal{B}}_{\Delta}(x^+_A,x_A^-){\mathcal{B}}_{\Delta}(x^+_B,x_B^-)\rangle \,.
\eea
In the case of spheres lying on the same hyperplane, the configuration space can be characterized by the single parameter $z$ which is related to usual conformal ratios via 
\bea\label{conf-parameters}
u\equiv \frac{|x_1-x_2|^2|x_3-x_4|^2}{|x_1-x_3|^2|x_2-x_4|^2}=z^2\,,\quad{\rm and }\quad  v=\frac{|x_1-x_4|^2|x_2-x_3|^2}{|x_1-x_3|^2|x_2-x_4|^2}=(1-z)^2\,.
\eea
After the identifications 
$x_1=x_A^-$, $x_2=x_A^+$  and $x_3=x_B^-$, $ x_4=x_B^+$, we get $z$ in terms of the geometric parameters $R_A$, $R_B$ and $L$ as
\bea\label{conformal-factor}
z=\frac{4R_AR_B}{L^2-(R_A-R_B)^2}\quad{\rm with }\quad z\in [0,1]\,,
\eea
where $L$ is the center-to-center distance between the spheres.
 
Using the integral representation (\ref{OPE-int-rep-6}), we can write down the following expression for the conformal block
\bea\label{CB-int}
G_{\Delta}(z)=\tilde{c}^2_\Delta \int_{D_A}\diff^d \xi \(\frac{|x_A^+-\xi ||x_A^--\xi |}{2R_A}\)^{\Delta-d} \!\! \int_{D_B}\diff^d \chi \(\frac{|x_B^+-\chi ||x_B^--\chi |}{2R_B}\)^{\Delta-d}\frac{1}{|\chi-\xi+L|^{2\Delta}}\,.\nonumber\\
\eea
From this expression, we can obtain an interesting result. Take $R_B\to 0$ while keeping $L$ and $R_A$ fixed. In that case, $G_{\Delta}(z)\sim z^\Delta$, $z\sim 4R_A R_B/|\chi-x_A^-||\chi-x_A^+|$ and we can pull the correlator outside the $\chi$ integral, and carry out the integral over $D_B$ using the normalization (\ref{normalizationOPE}). In this way, we get
\bea\label{rel-corr-1}
\(\frac{2R_A}{|x_A^+-\chi ||x_A^--\chi |}\)^{\Delta} =\tilde{c}_\Delta \int_{D_A}\diff^d \xi \(\frac{|x_A^+-\xi ||x_A^--\xi |}{2R_A}\)^{\Delta-d}\frac{1}{|\chi-\xi|^{2\Delta}}\,.
\eea
This equation is of pivotal importance in obtaining the main results of section \ref{MI-Local}. On the other hand, plugging (\ref{rel-corr-1}) into (\ref{CB-int}), one arrives at the following integral representation for the conformal block 
\bea\label{G-mink-rep}
G_{\Delta}(z)=\tilde{c}_\Delta \int_{D_A}\diff^d \xi \frac{\(2R_A\)^{d-\Delta}\(2R_B\)^{\Delta}}{|x_A^+-\xi |^{d-\Delta}|x_A^--\xi |^{d-\Delta}|L+x_B^+-\xi |^\Delta|L+x_B^--\xi |^\Delta} \,,
\eea
which is almost identical to the analog integral representation in the context of the Euclidean conformal blocks \cite{SimmonsDuffin:2012uy}. Note that for $d>\Delta>(d-2)/2$ (where the lower bound comes from unitarity) we can replace $\Delta\to d-\Delta$ in (\ref{rel-corr-1}), and obtain an equation that effectively represents the inverse of (\ref{rel-corr-1}). Such manipulation allows us to argue that the correlator $1/|\chi-\xi|^{2(d-\Delta)}$ acts as an inverse of $1/|\chi-\xi|^{2\Delta}$ in this particular case.

%\subsubsection{Closed formulas for conformal blocks}
%Some exact results for $d=2$ and $d=4$
%\bea
%G_{\Delta, l}^{(2d)}(u,v)&=&k_{\Delta+l}(z)k_{\Delta-l}(\bar{z})+k_{\Delta-l}(z)k_{\Delta+l}(\bar{z})  \nonumber\\
%G_{\Delta, l}^{(4d)}(u,v)&=&\frac{1}{(l+1)}\frac{z \bar{z}}{z -\bar{z}}\(k_{\Delta+l}(z)k_{\Delta-l-2}(\bar{z})-k_{\Delta-l-2}(z)k_{\Delta+l}(\bar{z})\)
%\eea
%where $u=z\bar{z}$, $v=(1-z)(1-\bar{z})$ and
%\bea
%k_{\beta}(x)\equiv x^{\beta/2} \,_2F_1\(\frac{\beta}{2},\frac{\beta}{2},\beta,x\)\,.
%\eea
%In the diagonal limit, this is when $z\to \bar{z}$ we have 
%\bea
%G_{\Delta, l}^{(2d)}(u,v)&=&2k_{\Delta+l}(z)k_{\Delta-l}(z) \nonumber\\
%G_{\Delta, l}^{(4d)}(u,v)&=&\frac{-2z^2}{(l+1)}\,k_{\Delta+l}(z) \partial_z k_{\Delta-l-2}(\bar{z})
%\eea

\subsection{Euclidean representation \label{euc-rep}}

As mentioned in subsection (\ref{mink-rep}) there is a Euclidean representation of conformal blocks that agrees almost exactly with (\ref{G-mink-rep}) \cite{SimmonsDuffin:2012uy}. That representation is constructed from distances among four Euclidean points as in the Lorentzian case. However, from the viewpoint of modular flows, we should be able to obtain a representation that involves the modular flows associated with spheres whose geometry in Euclidean signature is not determined by four coordinate points. In this section, we will construct such a representation. 

We will mimic the procedure followed in the Lorentzian context. An important hint follows from the structure behind the Lorentzian OPE representation (\ref{OPE-int-rep-6}). Namely, the role played by the conformal killing vector $|K|$. Such vector field can be analytically continued to the Euclidean setting where it is natural to conjecture that the following integral representation holds
\bea \label{OPE-Eucl}
\tilde{\mathcal{B}}_\Delta (|x_A|,R_A)=c_\Delta\int d^d x \[\frac{\(|\vec{x}|^2+\tau^2-R_A^2\)^2+4\tau^2 R_A^2}{(2R_A)^2}\]^{\frac{\Delta-d}{2}}\mathcal{O}_\Delta(x_A+x)\,,
\eea
where the Kernel in the above integral is given by $|K_E|^{\Delta-d}$. In this case, the integration domain is the whole space-time since the Euclidean modular flow of $A$ acts non-trivially everywhere. Following the same analysis as the Lorentzian case, one needs the above expression to be reduced to $(2R_A)^{\Delta}\mathcal{O}_\Delta(x_0)$ in the $R_A\to 0$ limit. This follows directly from the following delta function representation:
\begin{equation}
\label{eq:delta-representation}
\lim_{R\to 0} \frac{(2R)^{d-2 \Delta}}{\left((\left|\vec{x}\right|^2+\tau^2-R^2)^2+ 4 \tau^2 R^2\right)^{\frac{d- \Delta}{2}}} = \frac{1}{c_ \Delta} \delta^{(d-1)}(\vec{x}) \delta(\tau)\,,
\end{equation}
with
\begin{equation}
\label{Eucl-CB-coeff}
\frac{1}{c_ \Delta} = \frac{2^{2+d-2 \Delta} \pi^{\frac{d}{2}}\Gamma \left(  \frac{d}{2} \right) }{\Gamma \left( \frac{d+1}{2} \right) \Gamma \left( \frac{d-1}{2} \right) \Gamma \left( \frac{d- \Delta}{2} \right) }\int_0 ^\infty ds\; s^{d- \Delta} e^{2 s^2} K_{\frac{\Delta -1 }{2}}(2 s^2) \;_1 F_1 \left( \tfrac{1}{2}, \tfrac{d+1}{2}, -4 s^2 \right)\,. 
\end{equation}
 We use the fact that the limit in \eqref{eq:delta-representation} is zero assuming $d>2\Delta$, unless both $\tau$ and $\vec{x}$ equals zero, and  
\begin{equation}
\label{eq:limit_delta}
\lim_{R\to 0}\int_{\tau^2+\left|x\right|^2<\delta^2}d\tau d^{d-1}x \frac{(2R)^{d-2 \Delta}}{\left((\left|\vec{x}\right|^2+\tau^2-R^2)^2+ 4 \tau^2 R^2\right)^{\frac{d- \Delta}{2}}} = \frac{1}{c_ \Delta}
\end{equation}
for any $\delta>0$. Then, integrating this expression against an arbitrary function we can split the integral into a small region around the origin and the rest. For the integral around the origin, we can evaluate the arbitrary function at the origin and use \eqref{eq:limit_delta}, while the rest vanishes because the integrand will go to zero. We compute the Euclidean conformal block as
\bea\label{Eucl-CB}
\tilde{G}_{\Delta}(z)=\langle \tilde{\mathcal{B}}_\Delta (|x_A|,R_A)\tilde{\mathcal{B}}_\Delta (|x_B|,R_B) \rangle 
\eea
by direct analogy with the Lorentzian version. We can obtain the analogue of (\ref{rel-corr-1}) by taking the $R_B\to 0$ limit, the delta function representation (\ref{eq:delta-representation}) and $\tilde{G}_{\Delta}(z)\sim z^\Delta$. This is
\bea\label{Eucl-inverse-CB}
&&\frac{(2R_A)^\Delta }{\left[\(\(|\vec{x}_B|-R_A\)^2+\tau^2_B\) \(\(|\vec{x}_B|+R_A\)^2+\tau^2_B\)\right]^\frac{\Delta}{2}} \nonumber \\
&&\qquad \qquad \qquad \qquad \qquad =c_\Delta\int d^d x \[\frac{\(|\vec{x}|^2+\tau^2-R_A^2\)^2+4\tau^2 R_A^2}{(2R_A)^2}\]^{\frac{\Delta-d}{2}}\!\!\!\! \frac{1}{|x_B-x|^{2\Delta}}\,,
\eea
where for simplicity we chose $x_A=0$. We use (\ref{conf-parameters}) to define the conformal parameter $z$, and take the points $x_i$ to be the points at which the spheres $A$ and $B$ intersect the two-dimensional plane that contains their centers and the Euclidean time direction $\tau$. We can check that the above formula is consistent also with the $R_A\to 0$ limit.

The definition (\ref{Eucl-CB}) and relation (\ref{Eucl-inverse-CB}) allow us to re-write the conformal block in terms of the following Euclidean integral:
\bea\label{CB-Eucl}
\tilde{G}_{\Delta}(z)=c_\Delta \int\diff^d x \frac{\(2R_A\)^{d-\Delta}\(2R_B\)^{\Delta}}{\[\(|\vec{x}|^2+\tau^2-R_A^2\)^2+4\tau^2 R_A^2\]^{\frac{d-\Delta}{2}}\[\(|\vec{x}-\vec{L}|^2+\tau^2-R_B^2\)^2+4\tau^2 R_B^2 \]^{\frac{\Delta}{2}}} \,.
\eea
Note that since the result of the integral must be symmetric under exchange of $R_A$ and $R_B$, we deduce that the result of the integral is also invariant under $\Delta \to d-\Delta$, but this is not a property of the conformal block. Therefore, we need to project out one part of the Euclidean result, in line with \cite{SimmonsDuffin:2012uy}. Therefore, strictly speaking (\ref{OPE-Eucl}) is not the OPE block associated to the operator ${\cal O}_{\Delta}$ but an auxiliary object whose correlation function produces the wanted conformal block plus a shadow contribution.   

One might be worried that (\ref{CB-Eucl}) is inconsistent with the assumption $\tilde{G}_{\Delta}(z)\sim z^\Delta$ for small $z$, used around (\ref{Eucl-inverse-CB}), since from the above symmetry we get $\tilde{G}_{\Delta}(z)\sim a_1 z^\Delta +a_2\, z^{d-\Delta}$, for small $z$. However, one should recall that our delta function representation requires $d-2
\Delta>0$, which implies that the term $a_1z^\Delta$ dominates for small $z$. This saves our procedure from inconsistencies.

\section{Evaluation of determinant}

\subsection{Lorentzian \label{det-lorentzian}}
Using spherical coordinates for the point $f$ and the fact that under that sequence of coordinate transformations we are mapping a volume element in $\tilde{w}$ to a volume element in $f$, we get:
\bea
d^df&=&f_r^{d-2} \sin^{d-3}\(f_\psi\)df_t\wedge df_r \wedge df_\psi  \wedge d^{d-3} \Omega\, \nonumber\\
d^d\tilde{w}&=&\tilde{w}_r^{d-2} \sin^{d-3}\(\tilde{w}_\psi\)d\tilde{w}_t\wedge d\tilde{w}_r \wedge d\tilde{w}_\psi  \wedge d^{d-3} \Omega 
\eea
where we can identify the volume factor $d^{d-3} \Omega$ in both volume elements, thus we obtained:
\bea
d^df=\left|\det \(\frac{\partial f^\alpha}{\partial \tilde{w}^\beta} \)\right|d^d \tilde{w}\quad \to \quad \left|\det \(\frac{\partial f^\alpha}{\partial \tilde{w}^\beta} \)\right|=\frac{f_r^{d-3} \sin^{d-3}\(f_\psi\)}{\tilde{w}_r^{d-3} \sin^{d-3}\(\tilde{w}_\psi\)}\frac{1}{\tilde{w}_r}\left|\det\(\frac{\partial(f_t,f_x,f_z)}{\partial(\tilde{w}_t,\tilde{w}_r,\tilde{w}_\psi)}\)\right|\nonumber\\
\eea
where we also used the fact that $df_x\wedge df_z=f_r\,df_r\wedge df_\psi$ where $\{f_x,f_y\}$ are cartesian coordinates. 

Notice that the spherical symmetry is partially broken by the vector $\vec{L}=L\hat{z}$.
\bea 
f(\tilde{w})=x^J_A[\tilde{w},-s]-L-x^J_B\[\tilde{w}^L,-s\]
\eea
where $\tilde{w}^L=\tilde{w}-L$. In components
\bea
f_t(\tilde{w})&=& x^J_{t,A}[\tilde{w},-s]-x^J_{t,B}[\tilde{w}^L,-s]\nonumber\\
f_x(\tilde{w})&=&x^J_{r,A}[\tilde{w},-s]\sin\(\tilde{w}_\psi\)-x^J_{r,B}[\tilde{w}^L,-s]\sin\(\tilde{w}^L_\psi\)\nonumber\\
f_z(\tilde{w})&=& x^J_{r,A}[\tilde{w},-s]\cos\(\tilde{w}_\psi\)-L-x^J_{r,B}[\tilde{w}^L,-s]\cos\(\tilde{w}^L_\psi\)
\eea
where 
\bea
\tilde{w}^L_r=\sqrt{\tilde{w}_r^2+L^2-2L\tilde{w}_r\cos(\tilde{w}_\psi)}\quad{\rm and}\quad \sin(\tilde{w}^L_\psi)=\frac{\tilde{w}_r\sin(\tilde{w}_\psi)}{\sqrt{\tilde{w}_r^2+L^2-2L\tilde{w}_r\cos(\tilde{w}_\psi)}}
\eea
From these formulas we find a simple formula for the following combination of interest 
\bea
\frac{f_r^{d-3} \sin^{d-3}\(f_\psi\)}{\tilde{w}_r^{d-3} \sin^{d-3}\(\tilde{w}_\psi\)}=\(\frac{x^J_{r,A}[\tilde{w},-s]}{\tilde{w}_r}-\frac{x^J_{r,B}[\tilde{w}^L,-s]}{\sqrt{\tilde{w}_r^2+L^2-2L\tilde{w}_r\cos(\tilde{w}_\psi)}}\)^{d-3}
\eea
Evaluating the above expression on $\tilde{w}^*$, leads to
\bea
\frac{f_r^{d-3} \sin^{d-3}\(f_\psi\)}{\tilde{w}_r^{d-3} \sin^{d-3}\(\tilde{w}_\psi\)}\Bigg|_{\tilde{w}=\tilde{w}^*}=\frac{4^{d-3}\cosh^{d-3}\pi s\(\cosh^2\pi s-z\)^{\frac{d-3}2}}{z^{d-3}}
\eea
and 
\bea
\frac{1}{\tilde{w}_r}\left|\det\(\frac{\partial(f_t,f_x,f_z)}{\partial(\tilde{w}_t,\tilde{w}_r,\tilde{w}_\psi)}\)\right|_{\tilde{w}=\tilde{w}^*}=\frac{4^{3}\cosh^{3}\pi s\(\cosh^2\pi s-z\)^{\frac{3}2}}{z^{3}}
\eea
The final result for the determinant is thus:
\bea
\left|\det \(\frac{\partial f^\alpha}{\partial \tilde{w}^\beta} \)\right|=\frac{4^d\cosh^d\pi s\(\cosh^2\pi s-z\)^{\frac{d}2}}{z^{d}}
\eea
which is the quantity we were after.

\subsection{Euclidean \label{det-euclidean}}
Using spherical coordinates for the point $f$ and the fact that under that sequence of coordinate transformations we are mapping a volume element in $\tilde{w}$ to a volume element in $f$ we get:
\bea
d^df&=&f_r^{d-2} \sin^{d-3}\(f_\psi\)df_t\wedge df_r \wedge df_\psi  \wedge d^{d-3} \Omega\, \nonumber\\
d^d\tilde{w}&=&\tilde{w}_r^{d-2} \sin^{d-3}\(\tilde{w}_\psi\)d\tilde{w}_t\wedge d\tilde{w}_r \wedge d\tilde{w}_\psi  \wedge d^{d-3} \Omega 
\eea
where we can identify the volume factor $d^{d-3} \Omega$ in both volume elements, thus we obtained:
\bea
d^df=\left|\det \(\frac{\partial f^\alpha}{\partial \tilde{w}^\beta} \)\right|d^d \tilde{w}\quad \to \quad \left|\det \(\frac{\partial f^\alpha}{\partial \tilde{w}^\beta} \)\right|=\frac{f_r^{d-3} \sin^{d-3}\(f_\psi\)}{\tilde{w}_r^{d-3} \sin^{d-3}\(\tilde{w}_\psi\)}\frac{1}{\tilde{w}_r}\left|\det\(\frac{\partial(f_t,f_x,f_z)}{\partial(\tilde{w}_t,\tilde{w}_r,\tilde{w}_\psi)}\)\right|\nonumber\\
\eea
where we also used the fact that $df_x\wedge df_z=f_r\,df_r\wedge df_\psi$ where $\{f_x,f_y\}$ are cartesian coordinates. 

Notice that the spherical symmetry is partially broken by the vector $\vec{L}=L\hat{z}$.
\bea 
f(\tilde{w})=x_A[\tilde{w},-\theta]-L-x_B\[\tilde{w}^L,-\theta\]
\eea
where $\tilde{w}^L=\tilde{w}-L$. In components
\bea
f_t(\tilde{w})&=& x_t[\tilde{w},-\theta]-x_t[\tilde{w}^L,-\theta]\nonumber\\
f_x(\tilde{w})&=&x_r[\tilde{w},-\theta]\sin\(\tilde{w}_\psi\)-x_r[\tilde{w}^L,-\theta]\sin\(\tilde{w}^L_\psi\)\nonumber\\
f_z(\tilde{w})&=& x_r[\tilde{w},-\theta]\cos\(\tilde{w}_\psi\)-L-x_r[\tilde{w}^L,-\theta]\cos\(\tilde{w}^L_\psi\)
\eea
where 
\bea
\tilde{w}^L_r=\sqrt{\tilde{w}_r^2+L^2-2L\tilde{w}_r\cos(\tilde{w}_\psi)}\quad{\rm and}\quad \sin(\tilde{w}^L_\psi)=\frac{\tilde{w}_r\sin(\tilde{w}_\psi)}{\sqrt{\tilde{w}_r^2+L^2-2L\tilde{w}_r\cos(\tilde{w}_\psi)}}
\eea
From these formulas we find a simple formula for the following combination of interest 
\bea
\frac{f_r^{d-3} \sin^{d-3}\(f_\psi\)}{\tilde{w}_r^{d-3} \sin^{d-3}\(\tilde{w}_\psi\)}=\(\frac{x_r[\tilde{w},-\theta]}{\tilde{w}_r}-\frac{x_r[\tilde{w}^L,-\theta]}{\sqrt{\tilde{w}_r^2+L^2-2L\tilde{w}_r\cos(\tilde{w}_\psi)}}\)^{d-3}
\eea
Resulting in
\bea
\frac{f_r^{d-3} \sin^{d-3}\(f_\psi\)}{\tilde{w}_r^{d-3} \sin^{d-3}\(\tilde{w}_\psi\)}\Bigg|_{\tilde{w}=\tilde{w}^*_{1,\pm}}=0
\eea
\bea
\frac{f_r^{d-3} \sin^{d-3}\(f_\psi\)}{\tilde{w}_r^{d-3} \sin^{d-3}\(\tilde{w}_\psi\)}\Bigg|_{\tilde{w}=\tilde{w}^*_{2,\pm}}=\frac{4^{d-3}\sin^{d-3}\(\frac{\theta}{2}\)\(\sin^2\(\frac{\theta}{2}\)-z\)^{\frac{d-3}2}}{z^{d-3}}
\eea
and 
\bea
\frac{1}{\tilde{w}_r}\left|\det\(\frac{\partial(f_t,f_x,f_z)}{\partial(\tilde{w}_t,\tilde{w}_r,\tilde{w}_\psi)}\)\right|_{\tilde{w}=\tilde{w}^*_{1,\pm}}=\frac{4^{2}\sin^{2}\(\frac{\theta}{2}\)\(\sin^2\(\frac{\theta}{2}\)-z\)}{z^{2}}
\eea
\bea
\frac{1}{\tilde{w}_r}\left|\det\(\frac{\partial(f_t,f_x,f_z)}{\partial(\tilde{w}_t,\tilde{w}_r,\tilde{w}_\psi)}\)\right|_{\tilde{w}=\tilde{w}^*_{2,\pm}}=\frac{4^{3}\sin^{3}\(\frac{\theta}{2}\)\(\sin^2\(\frac{\theta}{2}\)-z\)^{\frac{3}2}}{z^{3}}
\eea
The final result for the determinant is thus:
\bea
\left|\det \(\frac{\partial f^\alpha}{\partial \tilde{w}^\beta} \)\right|=\frac{4^d\sin^d\(\frac{\theta}{2}\)\(\sin^2\(\frac{\theta}{2}\)-z\)^{\frac{d}2}}{z^{d}} \Theta\left(\sin^2 \tfrac{\theta}{2}-z\right)
\eea
for any of the roots.

\section{Resummation of the two-copy sector \label{resumation-2-copy}}

To understand how well the two-copy sector approximates the full mutual information, we explore the contribution to the total mutual information from our two-copy sector formula assuming a Cardy-like density of states. We focus on $d=2$, where we know the exact Cardy formula for the density of states at large $\Delta$. This asymptotic density of states  is given by
\bea
\rho(\Delta)=\(\frac{c}{96\Delta^3}\)^{1/4}e^{2\pi \sqrt{\frac{c\Delta}{6}}}\,.
\eea
We assume the theory only contains scalar operators. For $d=2$ we have
\bea\label{mutual-cardy-1}
I^{n \ell}_\Delta(A:B)&=&\frac{z^{2\Delta}}{64}\,\int_{0}^1 \frac{ w^{2\Delta}}{\sqrt{1-w}} \frac{\left(  1 + \sqrt{1-w\,z}\right)^{2(2-2\Delta)}}
{ \(1-w\,z \) }\,dw\,.
\eea
We want to compute the integral:
\bea
I(A:B)\approx \int d\Delta \rho(\Delta) I_\Delta(A,B)\,.
\eea
To do so, we find it convenient to re-express (\ref{mutual-cardy-1}) in terms of $\chi$ where $\chi=1-w z$. 
This leads to 
\bea
I^{n \ell}_\Delta(A:B)&=&\frac{1}{64 \sqrt{z}}\,\int_{1-z}^1 \frac{\(1+\sqrt{\chi}\)^4}
{ \sqrt{\chi-(1-z)} }\[\frac{\sqrt{1-\chi}}{1+\sqrt{\chi}}\]^{4\Delta}\,\frac{d\chi}{\chi}\,=\int_{1-z}^1 f(z,\chi)e^{-\frac{\Delta}{ \sigma^2(\chi)}}d\chi\,,
\eea
\bea
{\rm where}\quad \quad f(z,\chi)=\frac{1}{64 \sqrt{z}}\,\frac{\(1+\sqrt{\chi}\)^4}
{\chi \sqrt{\chi-(1-z)} }, \quad {\rm and}\quad \sigma^2(\chi)=\frac{1}{4 \log\(\frac{1+\sqrt{\chi}}{\sqrt{1-\chi}}\)}\,.
\eea
Thus, we can proceed to carry out the integral over $\Delta$. We start with
\bea
\int d\Delta \rho(\Delta)e^{-\frac{\Delta}{ \sigma^2(\chi)}} =
2\(\frac{c}{96}\)^{1/4} e^{\frac{\lambda_*^2}{\sigma^2(\chi)}}\int \frac{d\lambda}{\sqrt{\lambda}} e^{-\frac{\(\lambda-\lambda_*\)^2}{ \sigma^2(\chi)}}
\eea
where we define 
\bea
\lambda=\sqrt{\Delta}\,,\quad \lambda_*=\frac{\alpha \,\sigma^2(\chi)}{2}\,\qquad {\rm and }\qquad \alpha=2\pi \sqrt{\frac{c}{6}}\,.
\eea
We want to explore the result of the integral over large values of $\Delta$, or equivalently $\lambda$, where the Cardy formula applies. However, for $\lambda_*$ and $\sigma(\chi)$, satisfying $\lambda_*\gg 1$ and $\lambda_* \gg  \sigma(\chi)$ the unrestricted integral will be of order ${\cal O}(\sigma(\chi)/\sqrt{\lambda_*})$, which means 
\bea
\int d\Delta \rho(\Delta)e^{-\frac{\Delta}{ \sigma^2(\chi)}}\sim e^{\frac{\alpha^2\sigma^2(\chi)}{4}}\,,
\eea
and therefore, our approximation to the mutual information equals
\bea
I(A:B)&\approx &\int_{1-z}^1 f(z,\chi)e^{\frac{\alpha^2\sigma^2(\chi)}{4}}d\chi =\frac{1}{64\sqrt{z}}\,\int_{1-z}^1 \frac{\(1+\sqrt{\chi}\)^4}
{ \chi\sqrt{\chi-(1-z)} }e^{\frac{\pi^2 c}{24\log\(\frac{1+\sqrt{\chi}}{\sqrt{1-\chi}}\)}}d\chi\,.
\eea
Let us study the short-distance divergence in the above formula, this is the behavior as $z\to 1^-$. The lower-end contribution of the integral governs this limit, and can be approximated as follows:
\bea\label{I-resum-2-copy}
I(A:B)&\approx &\frac{\(1+\sqrt{1-z}\)^4}{64 \sqrt{z}\sqrt{1-z}}e^{\frac{\pi^2 c}{24\log\(\frac{1+\sqrt{1-z}}{\sqrt{z}}\)}}\,\int_{1-z}^1 \frac{d\chi}
{ \sqrt{\chi-(1-z)} }\nonumber\\
&=&\frac{2\(1+\sqrt{1-z}\)^4}{64 \sqrt{1-z}}e^{\frac{\pi^2 c}{24\log\(\frac{1+\sqrt{1-z}}{\sqrt{z}}\)}}\approx \frac{1}{16}\(\frac{R}{\delta}\) e^{\frac{\pi^2 c}{12}\(\frac{R}{\delta}\)}\,,
\eea
where we used $z\approx1-\delta^2/4R^2$.
The conditions $\lambda_*\gg 1$ and $\lambda_* \gg  \sigma(\chi)$ translates into $\sigma(\chi)\gg \sqrt{2/\alpha}$ and $\sigma(\chi)\gg 2/\alpha$. Since $\sigma(1-z)\to \infty$ as $z\to 1^-$, our conditions are fulfilled, and therefore, we can trust our approximated result. Thus, equation (\ref{I-resum-2-copy}) shows that the volume divergence (in $d=2$, a linear divergence) is exponentially enhanced when naively integrating our two-copy formula over the Cardy density of states. Therefore, we expect some of the contributions of the N-copy sector to compensate for this by being negative. 

\section{Constant and log terms in the two-copy sector\label{anomaly-coeffs}}
In this appendix, we present the explicit expression for the logarithmic and constant terms in the expansion of $I^{n \ell}_\Delta(A:B)$ at short distances. For even $d$ and odd $d$ these terms codify the contribution to $A^{(d)}$ and $F^{(d)}$ from $I^{n \ell}_\Delta(A:B)$, respectively. These results can be particularly interesting if $I^{n \ell}_\Delta(A:B)$ describes the mutual information of the GFF ${\cal O}_{\Delta}$ in some controlled approximation, as suggested by the approximate match with the results of \cite{Benedetti:2022aiw} explained in section \ref{UC-GFF}. We start from 
(\ref{I-hypers}) and split the sum over the hypergeometric functions labeled by $k$, into terms with $d-k=2n+ 1$ and with $d-k=2n$. Using the following expansion of hypergeometric functions around $z=1$,
\begin{align}
_2 F_1 \left(n+\tfrac{1}{2},1+2 \Delta, \tfrac{3}{2}+2 \Delta, 1-x\right) &= \frac{(n-1)! \Gamma \left( \frac{3}{2}+2 \Delta \right) }{\Gamma \left( \frac{1}{2}+n \right)  \Gamma \left( 1+2 \Delta \right) }\sum_{k=0}^{n-1} \frac{(\tfrac{1}{2})_k (1-n+2 \Delta)_k }{k! (1-n)_k} x^{k-n} \nonumber\\
&\!\!\!\!\!\!\!\!\!\!\!\!-\frac{(-1)^n \Gamma \left(  \frac{3}{2}+2  \Delta \right) }{\sqrt{\pi} n! \Gamma \left( 1-n+2 \Delta \right) } \left(\log(x) + H_{n- \frac{1}{2}}-H_n+H_{2 \Delta}\right)+ \mathcal{O}(x)\,,
\end{align}
and
\begin{align}
_2 F_1 \left(n,1+2 \Delta, \tfrac{3}{2}+2 \Delta, 1-x\right)&= \frac{\Gamma \left(  n-\frac{1}{2} \right)  \Gamma \left( \frac{3}{2}+2 \Delta \right) }{(n-1)!  \Gamma \left( 1+2 \Delta \right) }\sum_{k=0}^{n-1} \frac{(\tfrac{1}{2})_k (\frac{3}{2}-n+2 \Delta)_k }{k! (\frac{3}{2}-n)_k} x^{k-n+ \frac{1}{2}} \nonumber\\
&\qquad+\frac{\Gamma(\frac{1}{2}-n) \Gamma(\frac{3}{2}+2 \Delta)}{\sqrt{\pi} \Gamma(\frac{3}{2}-n+2 \Delta)}+ \mathcal{O}(\sqrt{x})\,,
\end{align}
where $(x)_k =(x)(x+1)\dots (x+k-1)$ is the Pochhammer symbol and $H_n$ are the harmonic numbers, we can directly identify the logarithmic term in $I^{n \ell}_\Delta(A:B)$ as
\begin{align}\label{log-I}
I_ {\Delta,\log}&(A:B)= -\frac{1}{4^{d+1}} \sum_{n=0}^{\lfloor \frac{d}{2}\rfloor} \frac{(-1)^n \Gamma(1+2d-4 \Delta) \Gamma(2\Delta +1) }{n!  \Gamma(d-2n) \Gamma(2+d+2n-4 \Delta) \Gamma(1-n+2 \Delta)}\log (1-z) \nonumber\\
&=-2\left(\frac{(2\Delta- d) \Gamma(\frac{1}{2}+d-\Delta)}{4 \Gamma(d+2)\Gamma( \frac{1}{2}-\Delta)}+\frac{ \Gamma(1+d-\Delta)}{2 \Gamma(d+2)\Gamma(-\Delta)}\right)
\log\(\frac{2R}{\delta}\) \stackrel{d=2}{=}- \frac{\Delta-1}{8}\log\(\frac{2R}{\delta}\)\,.
\end{align}
The above contribution for $d=2$ appears in the first sub-leading term for the mutual information at short distances presented in (\ref{short-d-2-copy-2d}). We find a logarithmic contribution for both even and odd numbers of dimensions in complete analogy with the mutual information of the GFF. However, to interpret the coefficient in (\ref{log-I}) as an anomalous coefficient $A^{(d)}$ we restrict to even $d$. In that case, we obtain
\bea\label{A-2copy-term}
A^{(d)}=\frac{(-)^{\frac{d}{2}}}{16}\left(\frac{(2\Delta- d) \Gamma(\frac{1}{2}+d-\Delta)}{\Gamma(d+2)\Gamma( \frac{1}{2}-\Delta)}+\frac{ 2\Gamma(1+d-\Delta)}{ \Gamma(d+2)\Gamma(-\Delta)}\right)\,.
\eea
As for the constant term, it receives contributions from all the hypergeometric functions and we obtain
\begin{align}\label{F-2copy-term}
I_ {\Delta,\text{const}}(A:B)&= \sum_{n=0}^{\lfloor \frac{d-1}{2}\rfloor} \left(\frac{4^{-d-1} \Gamma (2 \Delta +1) \Gamma (2 d-4 \Delta +1)\Gamma
   \left(-n-\frac{1}{2}\right)}{\Gamma (d-2 n-1) \Gamma \left(-n+2 \Delta +\frac{1}{2}\right) \Gamma (d+2
   n-4 \Delta +3)}\right.\nonumber \\
   &\left.-\frac{(-1)^n 4^{-d-1} \Gamma (2 \Delta +1) \Gamma (2 d-4 \Delta +1)\left(H_{2 \Delta }+H_{n-\frac{1}{2}}-H_n\right)}{n! \Gamma (d-2 n) \Gamma
   (-n+2 \Delta +1) \Gamma (d+2 n-4 \Delta +2)}\right)\nonumber\\
   &+\sum_{k=0}^\infty \frac{4^{-d-1} \Gamma (2 \Delta +1) \Gamma \left(\frac{k}{2}+\frac{1}{2}\right) \Gamma (2 (d-2 \Delta )+1)}{\Gamma (d+k+1) \Gamma \left(\frac{k}{2}+2 \Delta +\frac{3}{2}\right) \Gamma (-d-k+2 (d-2 \Delta )+1)}\\
&\!\!\!\!\!\!\!\!\!\!\!\!\!\!\!\!\!\!\!\!\!\!\!\!\!\!\!\!\!\!+\sum_{n=0}^{\lfloor \frac{d-1}{2}\rfloor}\sum_{k=0}^{n-1}\frac{4^{-d-1} \Gamma (2 \Delta +1) (-1)^{n-k} \Gamma \left(k+\frac{1}{2}\right) \Gamma (n) \Gamma (2 d-4
   \Delta +1)}{ \Gamma (k+1) \Gamma \left(n+\frac{1}{2}\right) \Gamma
   (d-2 n) (1-n)_k \Gamma (-k+n+1) \Gamma (-n+2 \Delta +1) \Gamma (d+2 n-4 \Delta +2)} \nonumber 
\end{align}
which does not take a simple expression even for $d=2$, the universal coefficient for odd values of $d$ would be $F^{(d)}=(-1)^{\frac{d-1}2} I_ {\Delta,\text{const}}(A:B)/2$. 

\bibliography{MI-MF}{}
\bibliographystyle{utphys}

\end{document}